\documentclass[aps,twocolumn,a4paper,showpacs]{revtex4}
\usepackage{graphicx}
\usepackage{amsmath}
\usepackage{amssymb}
\usepackage{enumerate}
\usepackage{subfigure}
\usepackage{tabularx}
\newcommand{\be}{\begin{equation}}
\newcommand{\ee}{\end{equation}}
\newcommand{\ben}{\begin{eqnarray}}
\newcommand{\een}{\end{eqnarray}}
\newcommand{\bes}{\begin{subequations}}
\newcommand{\ees}{\end{subequations}}

\newcommand{\bb}{\bibitem}
\newcommand{\sech}{{\rm sech}}

\newcommand{\vphi}{\varphi}
\begin{document}
\title{Exact solutions, energy and charge of stable Q-Balls}
\author{D. Bazeia$^{1}$, M.A. Marques$^1$, and R. Menezes$^{2,3}$}
\affiliation{$^1$Departamento de F\'\i sica, Universidade Federal da Para\'\i ba, 58051-970 Jo\~ao Pessoa, PB, Brazil}
\affiliation{$^2$Departamento de Ci\^encias Exatas, Universidade Federal da Para\'{\i}ba, 58297-000 Rio Tinto, PB, Brazil}
\affiliation{$^3$Departamento de F\'\i sica, Universidade Federal de Campina Grande, 58109-970, Campina Grande, PB, Brazil}
\begin{abstract}
In this work we deal with nontopological solutions of the Q-ball type in two spacetime dimensions. We study models of current interest, described by a Higgs-like and other, similar potentials which unveil the presence of exact solutions. We use the analytic results to investigate how to control the energy and charge to make the Q-balls stable. 
\end{abstract}
\date{\today}
\maketitle
\section{Introduction} 

In high energy physics, defect structures can engender topological or nontopological profile. Topological structures are stable thanks to topological arguments, because one can associate conserved currents to them, which are conserved by construction, due to the topological properties of the configurations \cite{b1}. However, nontopological structures \cite{b2,tdlee} do not attain topological charge to make them stable, but they can be stabilized in a diversity of cases, in particular  as Q-balls \cite{coleman,kolb}. 

The basic properties of Q-balls have been largely studied in the literature \cite{tdlee,coleman,kolb,41,41a,prl,42,kusenko,dm,q1,tuomas,minos,q2,sut,ku,sut1,ed,sta1,sta2}, and the investigations usually require a numerical approach, since it is hard to find analytical solutions for the nonlinear equations that govern the system. These methods allows us to understand basic properties \cite{tuomas} such as its existence and stability in certain (thin \cite{coleman} and thick wall \cite{kusenko}) limits, but this usually requires a fine tuning of parameters.

In the simplest case, the presence of Q-balls is related to the existence of global U(1) symmetries. However, in the Standard model the presence of global U(1) symmetries can be related to baryonic and leptonic charges, and in extended supersymmetric models, the scalar superpartners of baryons and leptons can condensate and give rise to Q-balls.
In this sense, Q-balls are of current interest to baryogenesis, for instance. As it is known, one can suggest that the baryon asymmetry of the universe appears via the Affleck-Dine mechanism \cite{ad}, as a feature of the flat direction inflation, with the flat direction condensate giving rise to Q-balls \cite{q1,q2}, which can latter decay under reheating \cite{rh}.

Numerical simulations were used to study stability under small fluctuations, interactions, and the scattering of Q-balls. We have searched the literature and found some exact, analytical solutions for Q-balls for Higgs-like potentials and other similar potentials, and this motivated us to investigate exact solutions for Q-balls, with focus on the study of properties such as shape, energy, charge, stability and splitting, without relying on numerical solutions.

We start the investigation in Sec.~\ref{sec:model}, where we describe the models and review some basic facts about Q-balls. We continue the study in Sec.~\ref{sec:illu}, where we investigate several specific models and study stability and other related features. We summarize the results and add some comments to end the work in Sec.~\ref{sec:end}. 

\section{The models}
\label{sec:model}

In order to investigate Q-balls, we consider the Lagrange density
\be
{\cal L} = \frac12 \partial_\mu{\bar\vphi} \partial^\mu \vphi - V(|\vphi|),
\ee
where $\vphi$ is a complex scalar field and
\be
V(|\vphi|) = \frac12 m^2\,|\vphi|^2-\frac13\alpha\,|\vphi|^{2+n}+\frac14 \beta\,|\vphi|^{2+2n},
\ee
is the potential, with $n=1,2,3\ldots$. This model was firstly considered in \cite{41a}, and for $n=1$ it reproduces the model investigated before in \cite{41,42}, and for $n=2$ it gives the model studied in \cite{tdlee}. Here, however, we explore the several values of $n$, and study the shape, energy, charge, stability and splitting of the corresponding Q-balls. In the potential, we consider $m>0$ as the mass parameter, with $\alpha$ and $\beta$ being real parameters. Using the rescaling
\be
\vphi \to \left(\frac{m^2}{\alpha}\right)^\frac{1}{n}\vphi, \quad x \to \frac{x}{m}, \quad {\cal L} \to m^2\left(\frac{m^2}{\alpha}\right)^\frac{2}{n} {\cal L},
\ee
we get the Lagrange density
\be\label{lgeneral}
{\cal L} = \frac12 \partial_\mu\vphi^* \partial^\mu \vphi - \frac12 |\vphi|^2 + \frac13|\vphi|^{2+n} - \frac14 a\,|\vphi|^{2+2n},
\ee
with $a=\beta m^2/\alpha^2$. We are working in $(1,1)$ spacetime dimensions, so the equation of motion has the form
\be
\ddot{\vphi} - \vphi^{\prime\prime} + \vphi - \frac{2+n}{3}\,|\vphi|^{n}\vphi + \frac{a (1+n)}{2}\,|\vphi|^{2n} \vphi = 0.
\ee
To search for Q-balls we take the usual ansatz
\be\label{ansatz}
\vphi(x,t)=\sigma(x)\,e^{i\omega t}.
\ee
The conserved Noether charge is
\be
Q = \frac{1}{2i} \int_\infty^\infty{dx\left({\bar\vphi}\dot{\vphi} - \vphi\dot{\bar\vphi}\right)},
\ee
or better
\be\label{charge}
Q= \omega \int_\infty^\infty{dx\, \sigma^2(x)}.
\ee
The equation of motion becomes 
\be\label{eom}
\sigma^{\prime\prime} = (1-\omega^2)\sigma -  \frac{2+n}{3}\,\sigma^{1+n} + \frac{a (1+n)}{2}\,\sigma^{1+2n}.
\ee
As usual, we consider the boundary conditions
\ben\label{bcond}
\sigma^\prime(0) = 0;\;\;\;\;\;\sigma(\infty) = 0.
\een
The above equation of motion \eqref{eom} can be seen in the form
\be\label{eqeff}
\sigma^{\prime\prime} = \frac{dU}{d\sigma},
\ee 
with $U=U(\sigma)$ being a kind of effective potential for the field $\sigma$. It has the form
\be\label{veffn}
U(\sigma) = \frac12 (1-\omega^2)\sigma^2 -\frac13 \sigma^{2+n} + \frac14 a\,\sigma^{2+2n}.
\ee

As one knows, in order to have solutions obeying the boundary conditions \eqref{bcond}, the effective potential $U(\sigma)$ has to have:
\begin{itemize}
\item symmetry breaking;
\item zeroes at points in which $\sigma$ is nonvanishing.
\end{itemize}
The first condition gives a superior bound for $\omega$, that is $\omega_+ = V^{\prime\prime}(0)=1$. The second condition gives the inferior bound $\omega_-=\sqrt{2V(\sigma_0)/\sigma_0^2}=\sqrt{1-2/(9a)}$, where $\sigma_0$ is the minimum of $V(\sigma)/\sigma^2$. Then, $\omega$ is such that
\be\label{condomega}
\omega_-<\omega<\omega_+.
\ee
We take $a\geq2/9$ to assure that $\omega$ is real. Thus, $\omega_-$ varies in the interval $[0,1)$, as $a$ increases from $2/9$ to larger and larger values. We then see that there is a large range of values for the parameter $a$ that can, in principle, give rise to Q-balls. In Fig.~\ref{fig1} we depict the potential for $a=4/9$ and $n=1$, for several values of
$\omega^2$ that satisfies the condition \eqref{condomega}; in the inset we illustrate how the potential vanishes near the origin for nonvanishing $\sigma$.  The equation of motion  \eqref{eom} admits the solution
\ben\label{soln}
\sigma(x) &=& \left(\frac{1-\omega^2}{2a}\right)^{\frac{1}{2n}}\left[\tanh\left(\frac{n\,\sqrt{1-\omega^2}}{2} \, x + b \right) \nonumber\right. \\
&&\left.-\tanh\left(\frac{n\,\sqrt{1-\omega^2}}{2}\, x - b \right)\right]^{\frac{1}{n}},
\een
where
\be\label{b}
b=\frac12\, \text{arctanh}\left({3\,\sqrt{\frac{(1-\omega^2)\,a}{2}}}\,\right).
\ee
Since $a\geq2/9$ and $\omega^2$ is bounded according to \eqref{condomega}, $b$ does not vanish, so the solution \eqref{soln} is bell-shaped, with amplitude given by the value of $\sigma$ that identifies a zero of the effective potential. We call this the point of return of the solution, $A$. It is given by
\be\label{amplituden}
A=\left[\frac{2-\sqrt{4-18a(1-\omega^2)}}{3a}\right]^\frac{1}{n},
\ee
and obeys $U(A)=0$. The point of return controls the amplitude of the solution, and it depends on $a$ and $\omega$, such that for a given $a$ it diminishes as $\omega^2$ increases. We define $A_-
\equiv \left[2/(3a)\right]^{1/n}$ as the limit of the amplitude in the case $\omega\to\omega_-$ and $A_+ \equiv 0$ as the limit of $\omega\to\omega_+$. Then, the amplitude of our solution is such that $A_+<A<A_-$. As $a\geq 2/9$, we have that the maximum amplitude starts at $A=3^{1/n}$ for $a=2/9$ and decreases up to zero as we increase $a$, for a given $n$. We define the width $L$ of the solution as the width at half height:
\be\label{widthn}
L\equiv \frac{4}{\sqrt{1-\omega^2}}\, \text{arcsech}\left(\sqrt{\frac{1-\tanh^2{b}}{2^n-\tanh^2{b}}} \,\right).
\ee 
We see from the above expression that the width of the solution increases as $\omega$ decreases toward $\omega_-$, making the solution to develop a plateau of height $A_-$.

It is possible to use the exact solution \eqref{soln} to calculate the charge from Eq.~\eqref{charge}; the general result has the form
\ben\label{chargen}
Q &=&\frac{2^{\frac{1}{n}-1}\sqrt{\pi}}{a^{\frac{1}{n}}}\frac{\omega\left(1-\omega^2\right)^{\frac{1}{n}-\frac12}}{2+n}\frac{\tanh^{\frac{2}{n}}{b}}{\sech^2{b}}\frac{\Gamma\left(2+\frac{2}{n}\right)}{\Gamma\left(\frac32+\frac{2}{n}\right)} \nonumber \\
&& \times \left[2(2+n)_2F_1\left(-\frac12,\frac{2}{n};\frac32+\frac{2}{n};\tanh^2 b\right) \right. \nonumber \\
&& \left. - n(1+\tanh^2{b})_2F_1\left(\frac12,\frac{2}{n};\frac32+\frac{2}{n};\tanh^2 b\right) \right], \nonumber \\
\een
where $_2F_1(a,b;c;z)$ is the Hypergeometric function and $\Gamma(z)$ is the Gamma function.

It is possible to obtain the behavior of the solution for $\omega\approx\omega_+$; it is
\be\label{sech}
\sigma_{+}(x) \approx A\,\sech^{\frac{2}{n}}\left(\frac{n\,\sqrt{1-\omega^2}}{2}\,x\right),
\ee
where $A$ is the amplitude, as it appears in Eq.~\eqref{amplituden}, which, for $\omega\approx\omega_+$, behaves as $A\approx \left[3\,(\omega_+^2-\omega^2)/2\right]^{1/n}$. We also have studied how the effective potential \eqref{veffn} behaves in the limit $\omega\to\omega_+$, when calculated with the solution \eqref{soln} at $x=0$, which is the amplitude \eqref{amplituden}; we define $\delta=\omega_+^2-\omega^2$ to get
\ben
\frac12(1-\omega^2)\sigma^2  &=& \frac12 \left(\frac{3\delta}{2} \right)^{\frac{2}{n}} \left[\delta + \frac{9a}{4n}\delta^2 + \mathcal{O}\left(\delta^3\right)\right], \\
\frac13\sigma^{2+n} &=& \frac12\left(\frac{3\delta}{2} \right)^{\frac{2}{n}} \left[\delta + \frac{9a(2+n)}{8n}\delta^2 + \mathcal{O}\left(\delta^3\right)\right], \nonumber\\
\\
\frac14 a\,\sigma^{2+2n} &=&\frac{9}{16}\left(\frac{3\delta}{2} \right)^{\frac{2}{n}} \left[a\delta^2 + \mathcal{O}\left(\delta^3\right)\right].
\een
We see that the terms with $\sigma^2$ and $\sigma^{2+n}$ are proportional to $\delta^{1+2/n}$ and the term with $\sigma^{2+2n}$ is proportional to $\delta^{2+2/n}$. Then, when $\omega\to\omega_+$ we have $\delta\to0$, and the term $\sigma^{2+2n}$ can be neglected and the solution in this aproximation is given by Eq.~\eqref{sech}; this result fits within the thick wall approximation, as it is considered in \cite{kusenko} for small Q-Balls. In this regime, for the charge \eqref{chargen}, we have:
\ben\label{qp}
Q_+ &=& \frac{\sqrt{\pi}}{2(2+n)}\frac{\Gamma\left(2+\frac2n\right)}{\Gamma\left(\frac32+\frac2n\right)}\left(\frac32\right)^{\frac2n} \delta^{\frac2n -\frac12} \nonumber\\
&&\times \left[4+n +\frac{9a(2+n)-n(4+n)}{2n} \delta +\mathcal{O}\left(\delta^{\frac32}\right)\right]. \nonumber \\
\een
Then, when $\omega\to\omega_+$, the charge tends to zero for $n<4$. For $n=4$, it tends to a positive constant. For $n>4$, it diverges, and this modifies the stability of the system. The width \eqref{widthn} behaves as
\be
L_+ = 4 \, \textrm{arccosh}\left( 2^{\frac{n}{2}} \right)\delta^{-\frac12} +\frac{9a\sqrt{1-2^{-n}}}{4}\delta^{\frac12}+ \mathcal{O}\left(\delta^{\frac32}\right),
\ee
which, in the limit $\omega\to\omega_+$, diverges faster as we increase $n$. The amplitude \eqref{amplituden} behaves as
\be
A_+= \left(\frac{3\delta}{2}\right)^{\frac1n} \left[ 1+ \frac{9a}{8n} \delta + \mathcal{O}\left(\delta^{2}\right) \right]
\ee
Then, as $\omega$ approaches $\omega_+$, we see that the width becomes increasingly larger, whist the amplitude becomes smaller and smaller.

We see from \eqref{b} that $b$ increases to larger and larger values, as $\omega$ decreases toward its lower bound, $\omega_-$, which depends on $a$. Thus, one can rewrite the width \eqref{widthn} in terms of the parameters $n$, $a$ and $b$, to study its asymptotic behavior with respect to $b$, for a given $a$ and $n$; we can write
\be\label{asymptwidth}
L_{-} = 6\sqrt{2a} \left[\frac12 \ln\left(2^n-1\right) + b + \frac{e^{-2b}}{2^n-1} + {\cal O}\left(e^{-4b}\right)\right].
\ee
Then, we see that the width increases linearly with $b$, as $b$ increases to larger values. We can implement a similar investigation, to show that the charge \eqref{charge} can be written as 
\ben
Q_{-} &=& \left(\frac{2}{3a}\right)^{\frac{2}{n}}\frac{\sqrt{18a-4}}{n} \nonumber \\
&\times & \left[\frac{n}{2} - H\!_{\frac2n} +2b -2e^{2b}\right] \ +{\cal O}\left(e^{-4b}\right), \nonumber \\
\een
which shows a behavior similar to the width. In this case, $H_m$ is the $m^{th}$ Harmonic Number. Thus, both the charge and the width vary linearly with $b$, for larger values of $b$.

The model under investigation has the energy-momentum tensor
\be\label{emt}
T_{\mu\nu} = \frac12\partial_\mu{\bar\vphi}\partial_\nu\vphi + \frac12\partial_\mu\vphi\partial_\nu{\bar\vphi} - \eta_{\mu\nu}{\mathcal L}.
\ee
The energy density can be calculated from \eqref{emt} with the Lagrange density \eqref{lgeneral}. It is given by
\be
\epsilon= \epsilon_k + \epsilon_g + \epsilon_p  ,
\ee
where 
\ben
\epsilon_k  &=& \frac12|\dot{\vphi}|^2, \\
\epsilon_g &=& \frac12|\vphi^{\prime}|^2, \\
\epsilon_p &=&  \frac12 |\vphi|^2 - \frac13|\vphi|^{2+n} + \frac14 a\,|\vphi|^{2+2n},
\een
are the kinetic, gradient and potential portions of the energy density. After using the ansatz \eqref{ansatz}, the energy density becomes
\be\label{energydens1}
\epsilon = \frac12{\sigma^\prime}^2 + \frac{\omega^2+1}{2}\sigma^2 - \frac13\sigma^{2+n} + \frac14 a\,\sigma^{2+2n}.
\ee
We can substitute Eq.~\eqref{soln} in the above equation to get the explicit form of the energy density; however, the full expression  is cumbersome, and so we omit it here. We have done a closer inspection on the energy density, searching for any possible change of behavior. For a given $a$, one can show that it starts to split, the splitting appearing for $\omega$ in the interval $\omega_-<\omega<\omega_c$, where
\be\label{omegac}
\omega_c= \sqrt{\frac{n}{a}}\frac {{\left(\!\sqrt{(n\!+\!2)^2\!\!-\!18a(n\!+\!1)}\!+\!(9a\!-\!1)(n\!+\!1)\!-\!1\right)}^{1/2}}{3\,(n+1)}
\ee
Since $\omega_c$ must be real, we see that the splitting appears if 
\be\label{ac}
\frac29 \leq a < a_c=\frac{1}{18}\frac{(n+2)^2}{n+1}.
\ee
The other components of the energy-momentum tensor \eqref{emt} are
\ben
T_{01} &=& Re\left(\dot{\bar\vphi} \vphi^\prime\right), \\
T_{11} &=& \frac12|\vphi^\prime|^2 + \frac12|\dot{\vphi}|^2 - \frac12 |\vphi|^2 + \frac13|\vphi|^{2+n} - \frac14 a|\vphi|^{2+2n}. \nonumber \\
\een
With the ansatz \eqref{ansatz}, they become
\ben
T_{01} &=& 0, \\
T_{11} &=& \frac12{\sigma^\prime}^2 - \frac{1-\omega^2}{2} \sigma^2 + \frac13\sigma^{2+n} - \frac14 a\,\sigma^{2+2n}.
\een
Since the energy-momentum tensor is conserved, i.e., $\partial_\mu T^{\mu\nu}=0$, we see that $T_{11}$ is constant. Although $T_{11}$ is not zero in general, we see that it vanishes on shell, that is, if we take for $\sigma$ the solution \eqref{soln}. We also note that the equation of motion \eqref{eqeff} can be written in first-order form, compatible with $T_{11}=0$, as
\be  
\sigma^{\prime2}=\left(1-\omega^2-\frac23 \sigma^n+\frac12 a\, \sigma^{2n}\right) \sigma^2,
\ee
and it is solved by the solution \eqref{soln}. Furthermore, the condition $T_{11}=0$ lead us to 
\be\label{kgp}
\epsilon_p = \epsilon_k + \epsilon_g.
\ee
This can be integrated to give $E_p = E_k+E_g$, similar to the virial theorem. The same expression can be obtained from scaling arguments \cite{dt,dt1}. However, the condition that appears from the energy densities is stronger, because it is valid locally,
for any $x$. 

We can integrate the energy density to find the total energy
\be
E = 2\,(E_k + E_g).
\ee
The potential energy does not appear in the above expression because of the constraint given by Eq.~\eqref{kgp}. We note that the kinetic energy can be written in terms of the charge \eqref{chargen}
\be\label{ken}
E_k = \frac{\omega\, Q}{2}.
\ee
The gradient energy is given by
\ben\label{gen}
E_g &=& \frac{\sqrt{\pi}\,2^{\frac{1}{n} -1}}{n\,a^{1/n}}\frac{\Gamma\left(\frac{2}{n}\right)}{\Gamma\left(\frac32 + \frac{2}{n}\right)}  \left( 1-\omega^2 \right)^{\frac{1}{n}+\frac12} \tanh^\frac{2}{n}{b} \nonumber \\
&&\times _2F_1\left(-\frac12,\frac{2}{n};\frac32+\frac{2}{n};\tanh^2{b}\right).
\een

Let us now turn attention to the stability of the Q-ball; see, e.g., Refs.~\cite{coleman,sta1,sta2} and references therein. To make the investigation complete, we highlight the possibilities: 
\begin{itemize}
\item Quantum mechanical stability, which concerns stability against decay into free particles. As it was stated in Eq.~\eqref{condomega}, Q-Balls solutions exist for $\omega$ in a specific range of values. The Q-ball is stable if the ratio between the energy and the charge satisfies $E/Q < \omega_+$.
\item Classical stability, which concerns stability under small perturbations of the field configuration. The Q-ball is classically stable
if $dQ/d\omega<0$. This means that the charge $Q$ is monotonically decreasing with $\omega$.
\end{itemize}
There is another type of stability, against fission, which requires that $d^2E/dQ^2<0$. However, as we know that $\partial E/\partial Q=\omega$, it is straightforward to show that classically stable Q-balls are also stable against fission. In our model, the equation \eqref{qp} allows us to see that the charge is infinity for $\omega\to\omega_+$, when $n>4$. Thus, models with $n>4$ are classically unstable. In the next section we study some particular cases of the general potential \eqref{veffn}. 

\section{Illustration}
\label{sec:illu}

Let us now illustrate our results, investigating several distinct possibilities, controlled by the integer $n$, which we choose to be $n=1$, $n=2$, and $n=3$. 

\subsection{The case $n=1$}
\label{sec:n=1}

\begin{figure}[t!]
\includegraphics[width=6.0cm]{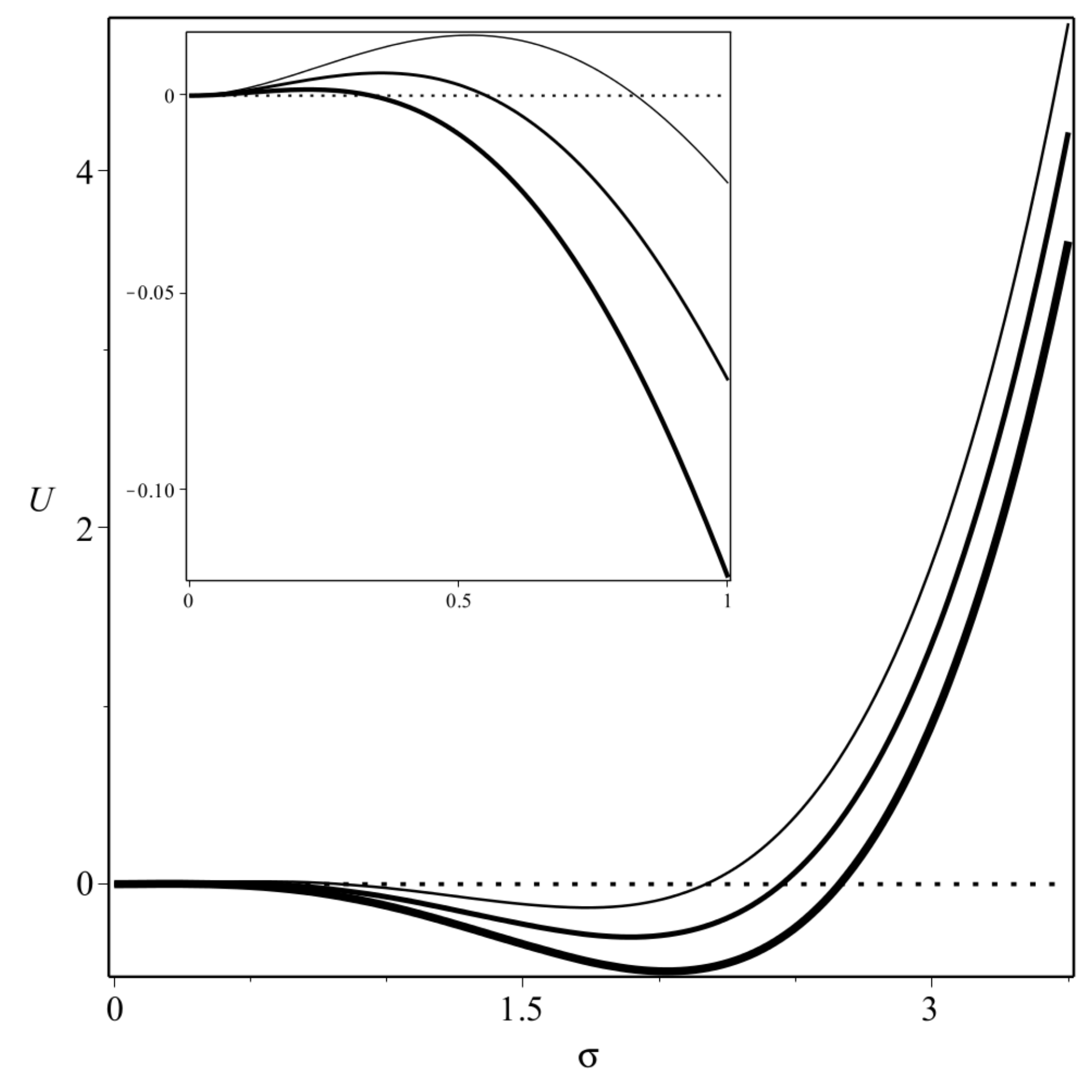}
\caption{The effective potential \eqref{veffn} depicted for $n=1$, $a=4/9$ and $\omega^2=0.6,0.7$ and $0.8$. The thickness of the line increases with $\omega^2$. In the inset, one shows the behavior of the effective potential for $\sigma\in [0,1]$.}\label{fig1}
\end{figure}

We take $n=1$ in Eq.~\eqref{veffn} to get
\be\label{veff1}
U(\sigma) = \frac12 (1-\omega^2)\sigma^2 -\frac13 \sigma^{3} + \frac14 a\,\sigma^{4}.
\ee
This potential contains up to the fourth-order power in the scalar field, so it is of current interest to high energy phenomenology. It was studied before in \cite{41,42}, but here we go further to add some new features, not seen before. The potential is plotted in Fig.~\ref{fig1}.

\begin{figure}[t!]
\includegraphics[width=4.20cm]{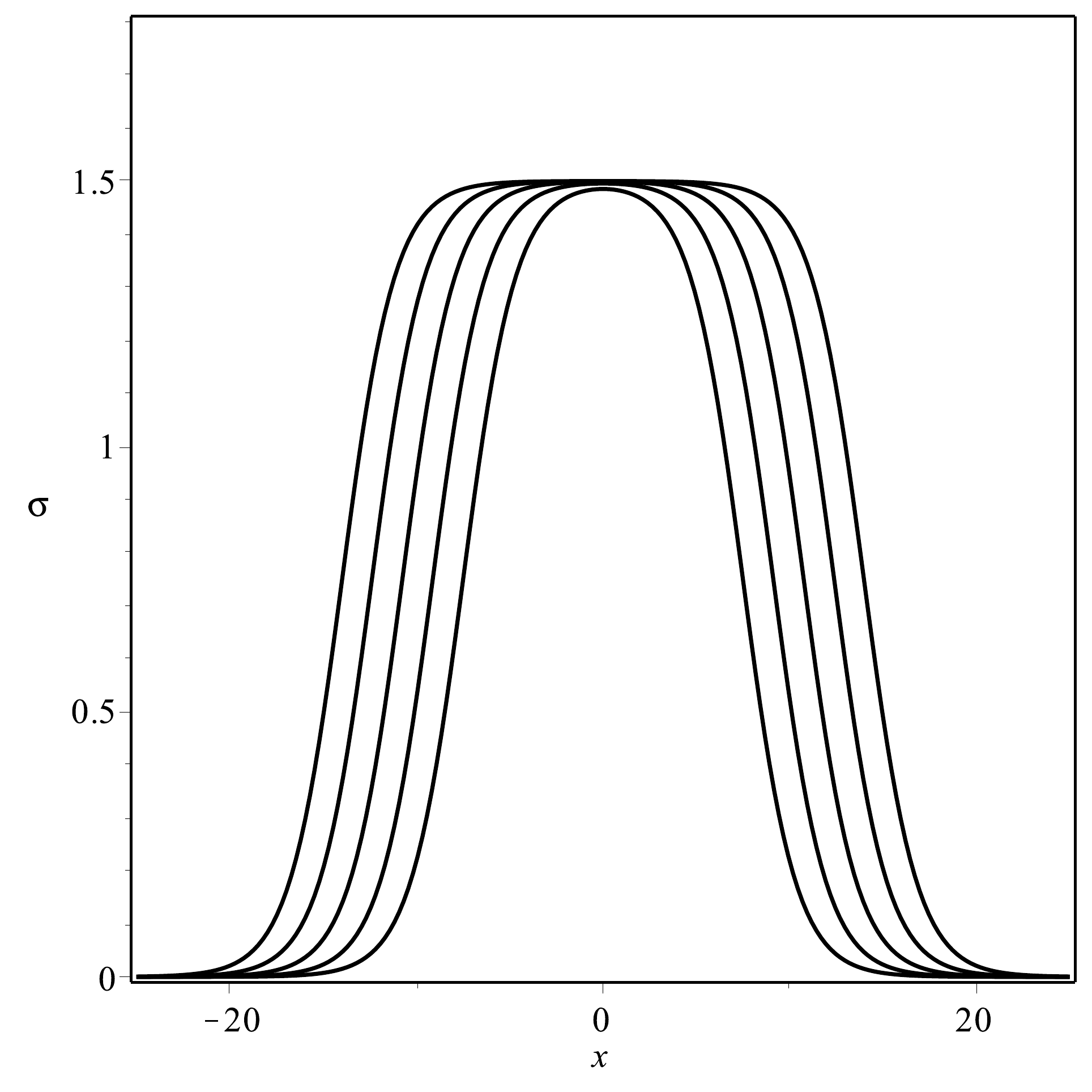}
\includegraphics[width=4.24cm]{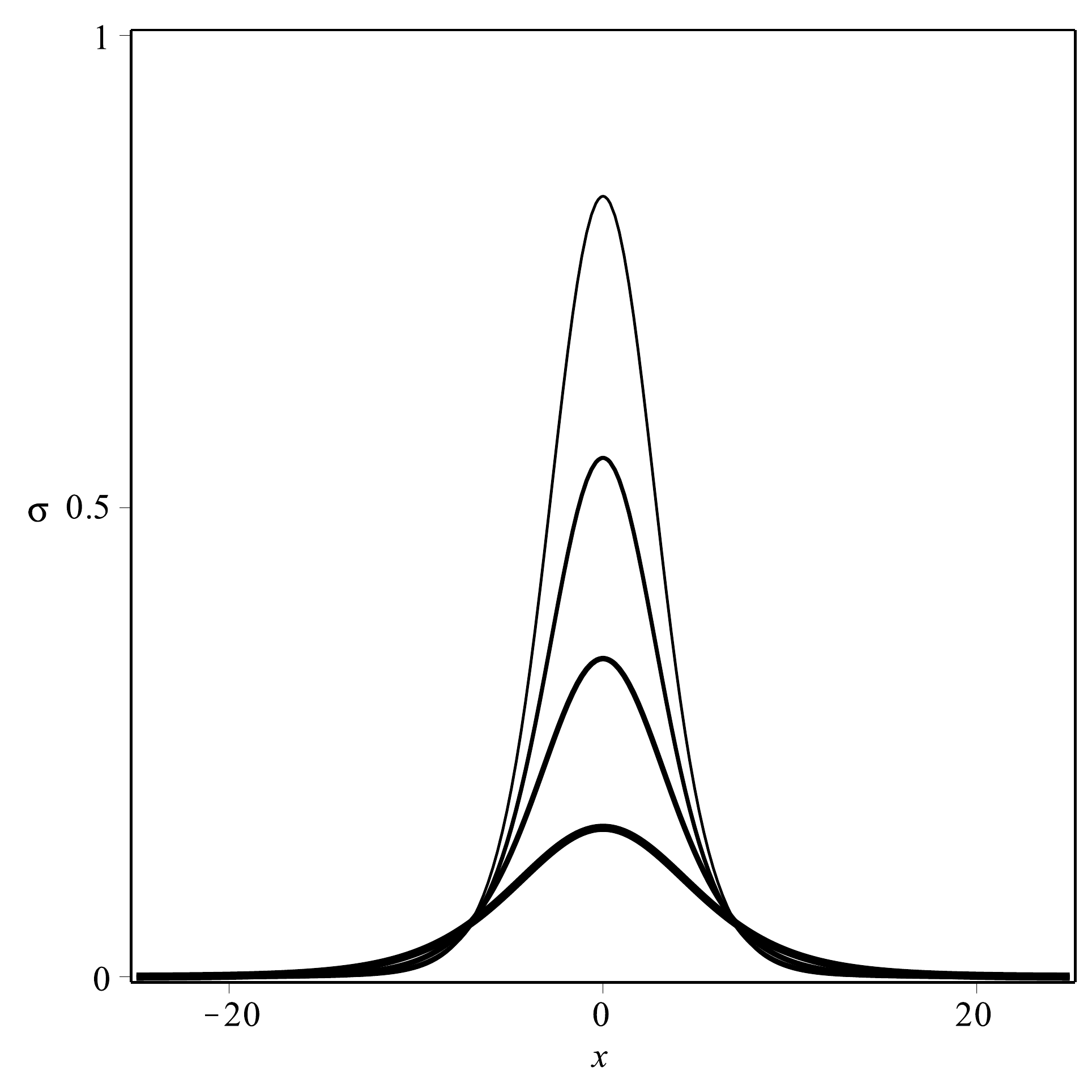}
\caption{The solution \eqref{soln} depicted for $n=1$ and $a=4/9$, with $\omega^2=0.5+5\epsilon$, $\epsilon=10^{-9},10^{-8},10^{-7},10^{-6}$ and $10^{-5}$ (left), and with $\omega^2=0.6, 0.7,0.8$ {\rm and} $0.9$ (right). The plateau in the left panel increases as $\omega$ approaches $\omega_-$. The thickness of the line in the right panel increases as $\omega^2$ increases.}\label{fig2}
\end{figure}

The solution can be found setting $n=1$ in Eq.~\eqref{soln}, which is depicted in Fig.~\ref{fig2} for $a=4/9$ ($\omega^2_-=0.5$), for several values of $\omega$ obeying Eq.~\eqref{condomega}. In Fig.~\ref{fig2}, in the left panel we can see the plateau for $\omega\approx\omega_-$ and in the right panel it is shown that the amplitude of the solution decreases as $\omega$ increases toward $\omega_+$.
 
\begin{figure}[t!]
\includegraphics[width=5.4cm]{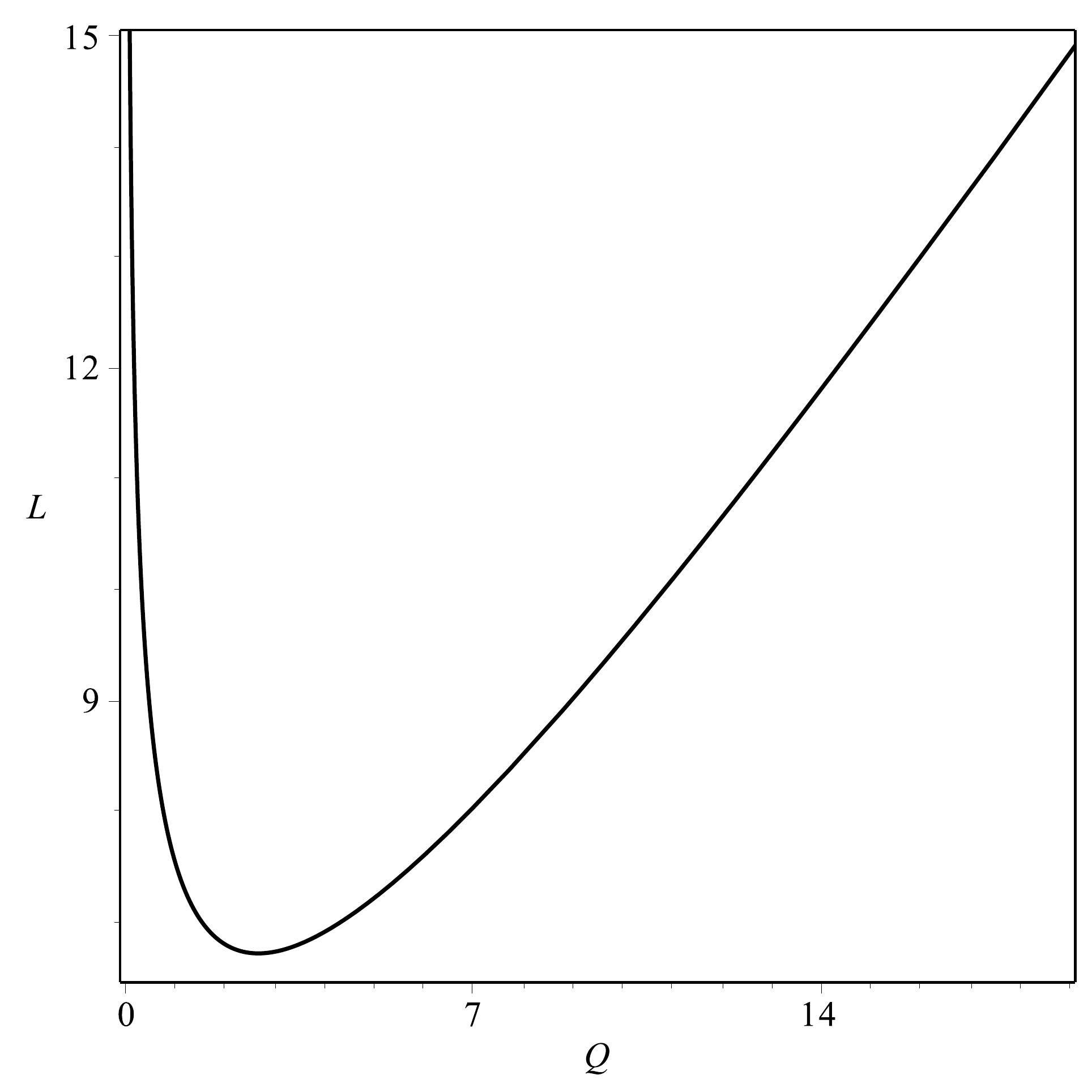}
\caption{The behavior of the width \eqref{widthn} for $n=1$ as a function of the charge \eqref{charge1} for $a=4/9$.}\label{chargewidth1}
\end{figure}

The charge \eqref{chargen} simplifies for $n=1$, becoming
\be\label{charge1}
Q=\frac{4\omega\sqrt{1-\omega^2}}{a} \left(2b\coth(2b)-1\right).
\ee
We see that $Q\to0$ as $\omega\to\omega_+$, for any $a$. Also, specifically for $a=2/9$ we have $\omega_-=0$, which makes $Q\to0$ for
$\omega\to\omega_-=0$. For $a>2/9$ we have $Q\to\infty$ for $\omega\to\omega_-$. The width can be easily obtained from Eq.~\eqref{widthn}. In Fig.~\ref{chargewidth1}, we display the width as a function of the charge for $a=4/9$. The minimum of this curve can be calculated numerically for each value of $a$. In particular, for $a=4/9$, the minimum appears for $\omega_m \approx 0.7757294$, $Q \approx 2.6728904$ and $L\approx 6.7119190$. We define this as the point that separates small Q-balls from large Q-balls.

The kinetic \eqref{ken} and gradient \eqref{gen} energies simplify to
\ben
E_k \!\!&=&\!\! \frac{2\omega^2\sqrt{1-\omega^2}}{a} \left(2b\coth(2b)-1\right)\\
E_g\!\!&=&\!\! \frac{(1-\omega^2)^{3/2}}{3a} \nonumber \\
\!\!& \times &\!\! \left(\frac{1+3(8b+3)e^{4b} + 3(8b-3)e^{8b} - e^{12b}}{\left(1 - e^{4b}\right)^3}\right)\!.\;\;
\een

\begin{figure}[htb!]
\includegraphics[width=5.4cm]{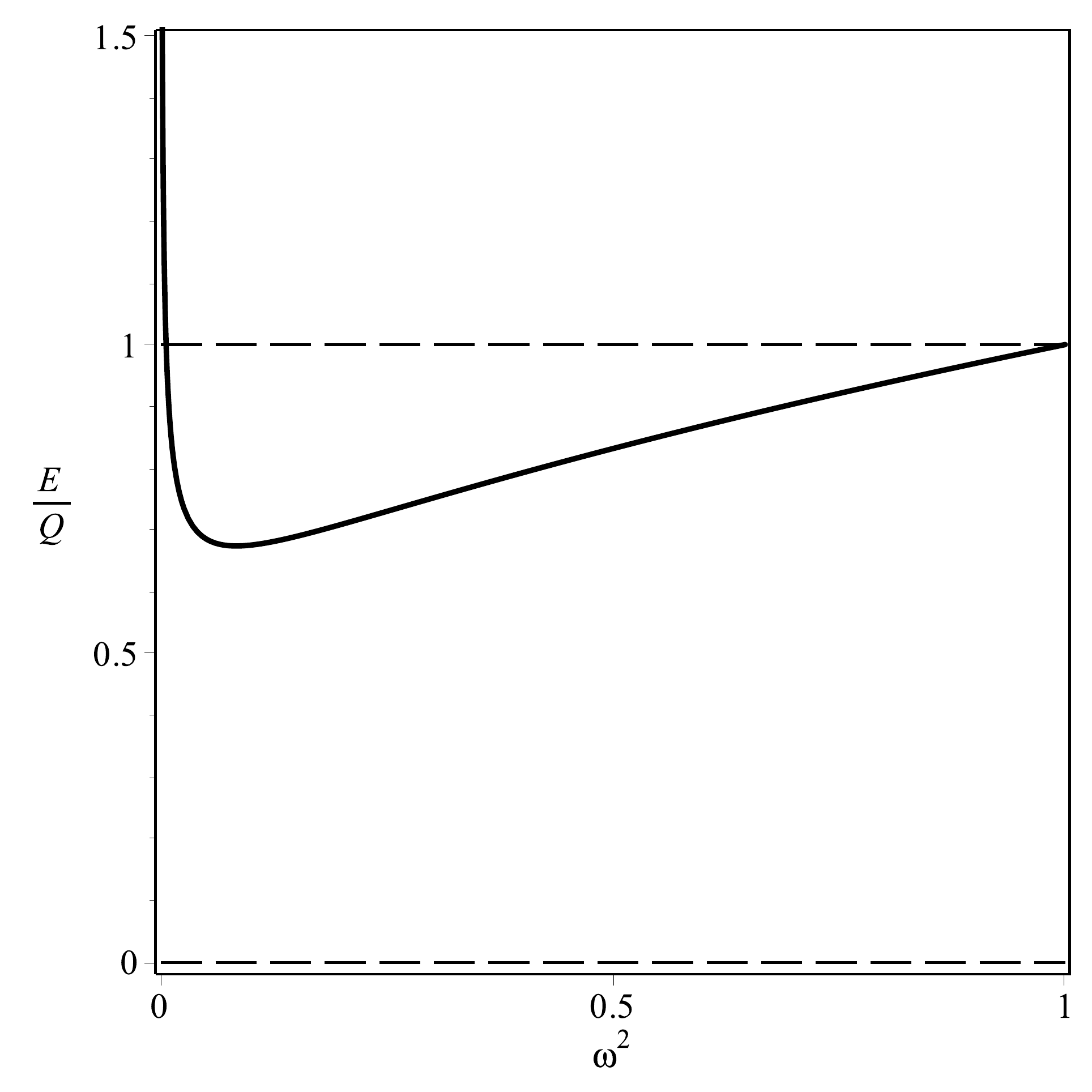}\\
\includegraphics[width=5.4cm]{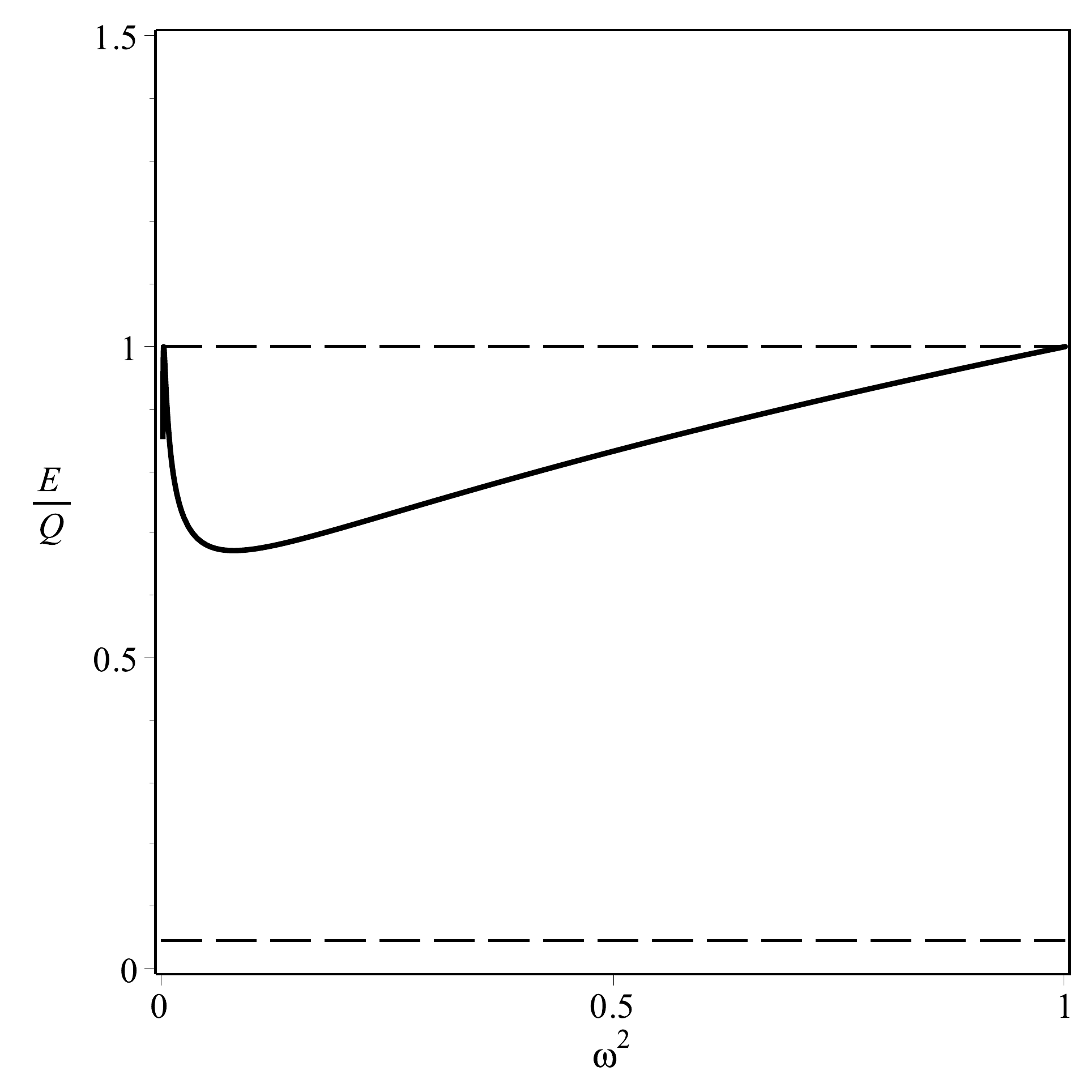}\\
\includegraphics[width=5.4cm]{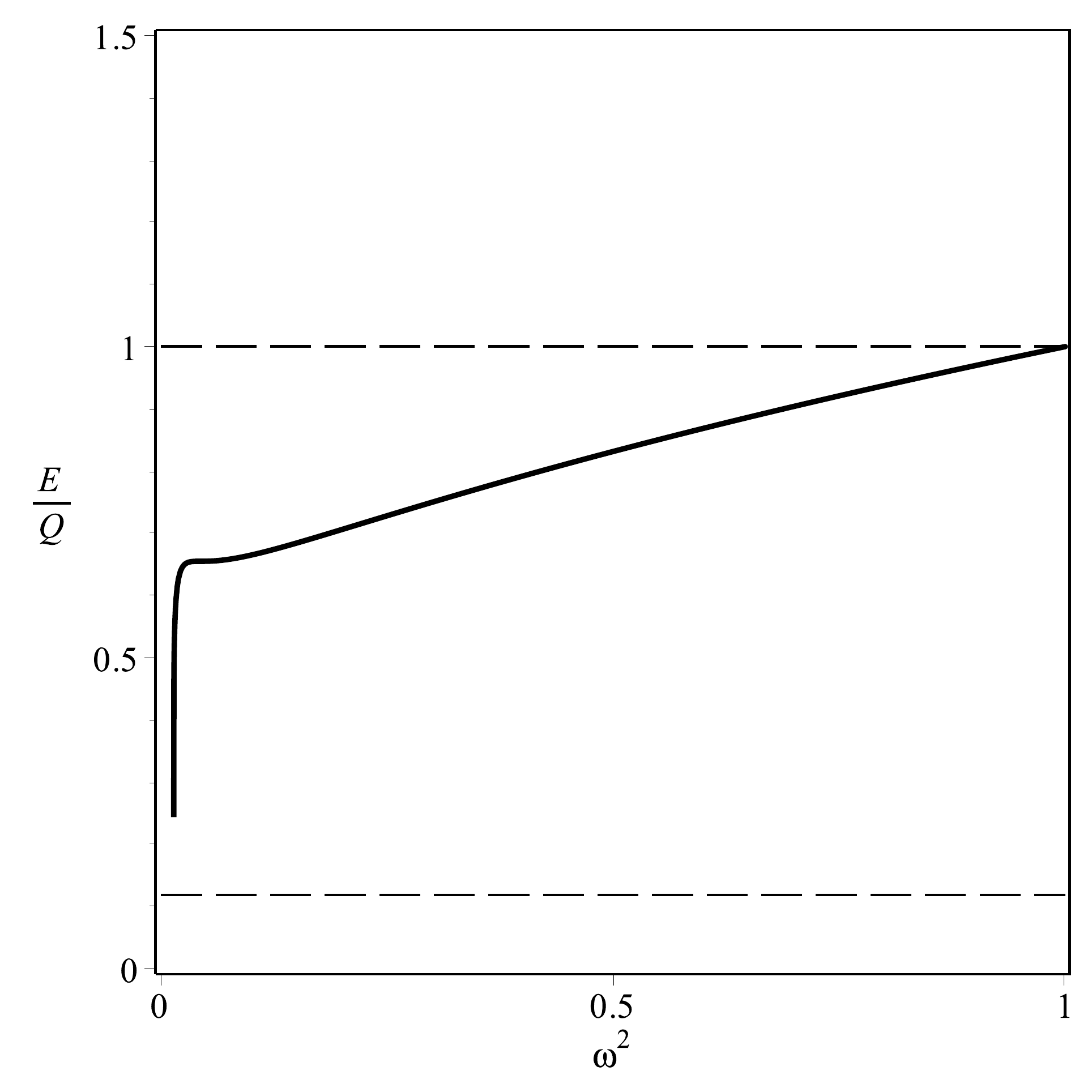}
\caption{The ratio $E/Q$ as a function of $\omega^2$, for the parameter $a$ as $a_0$ (top panel), $a_1$
(center panel) and $a_2$ (bottom panel). The region in between the two dashed horizontal lines assures quantum mechanical stability of the Q-Ball.}\label{fig4}
\end{figure}

\begin{figure}[t!]
\includegraphics[width=5.2cm]{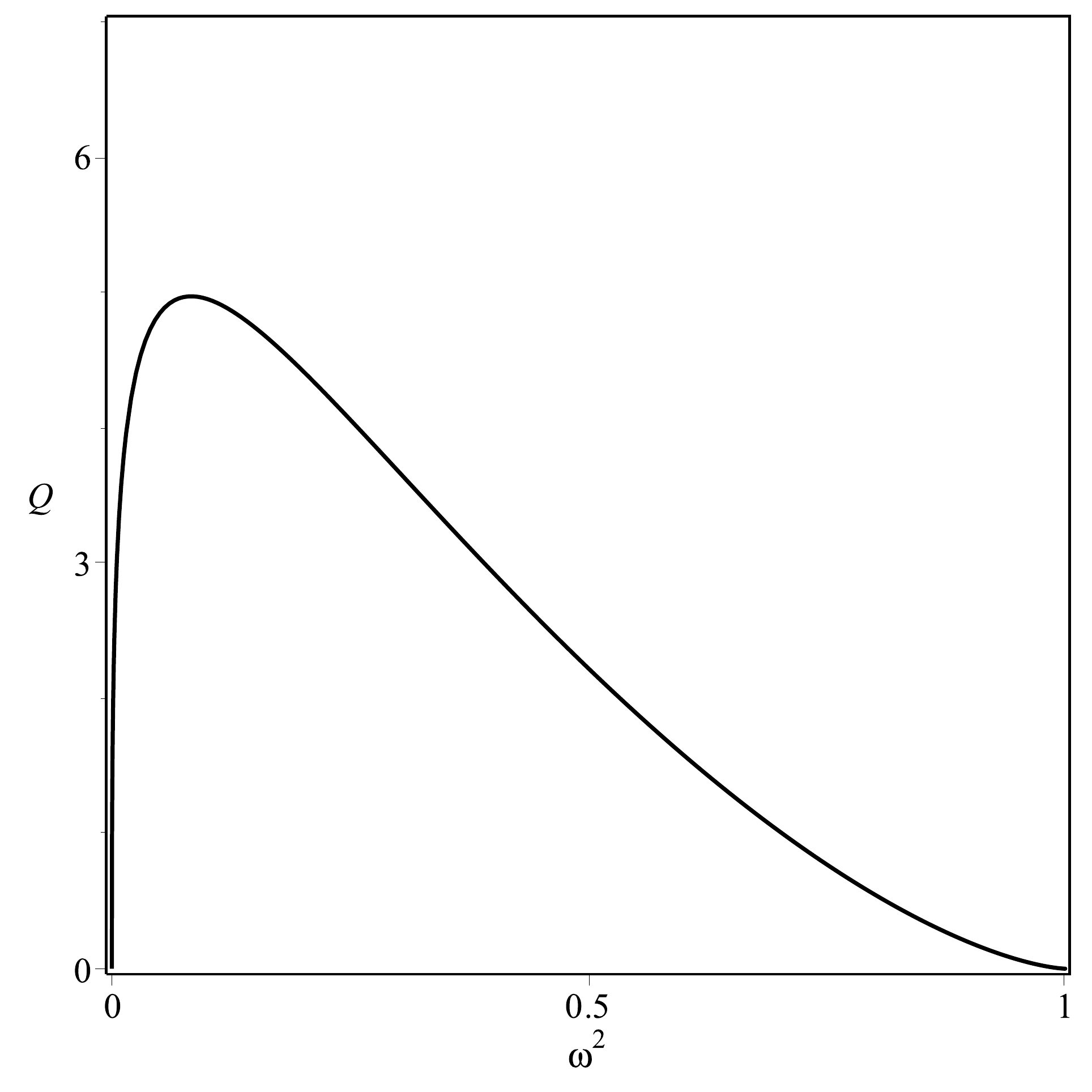}
\includegraphics[width=5.2cm]{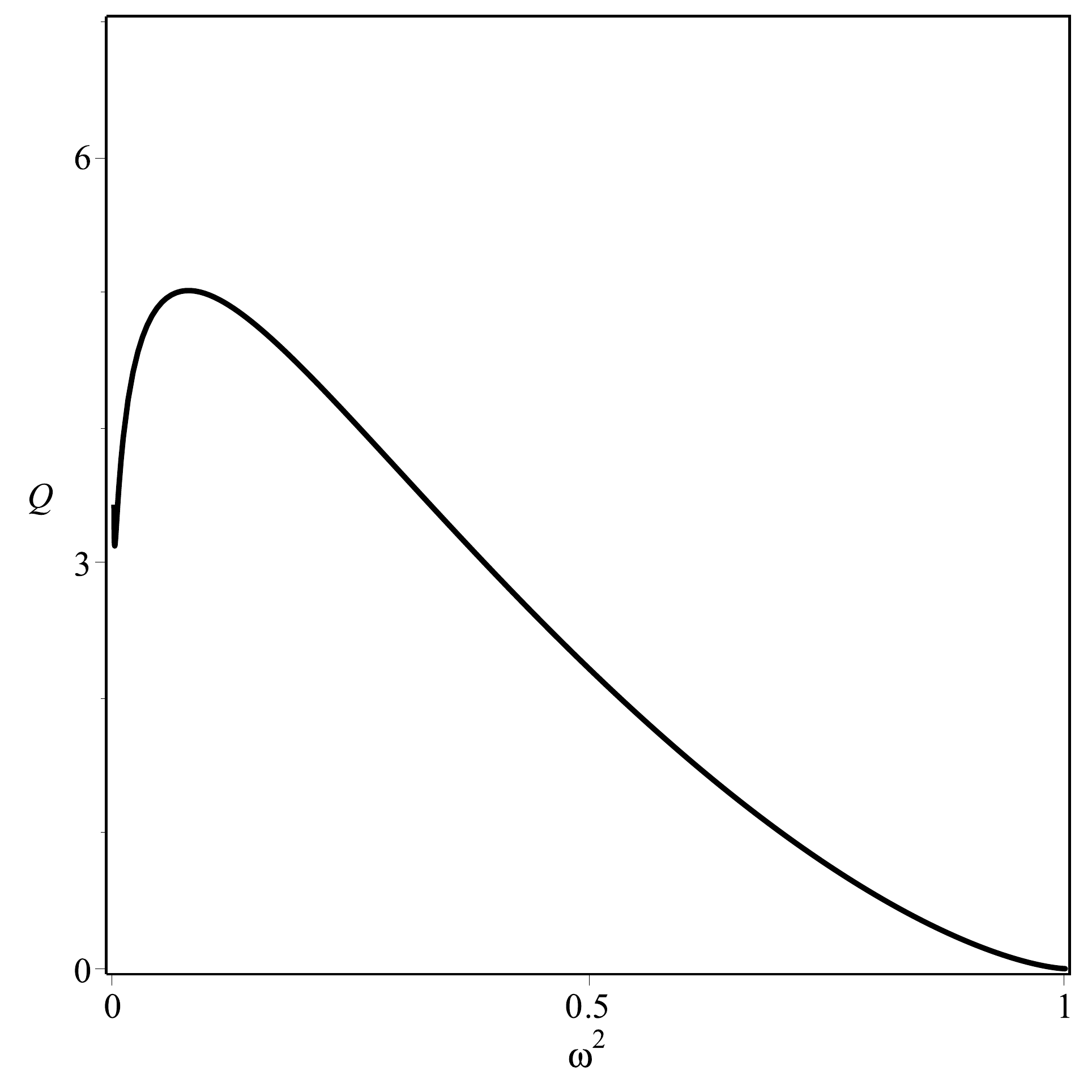}
\includegraphics[width=5.2cm]{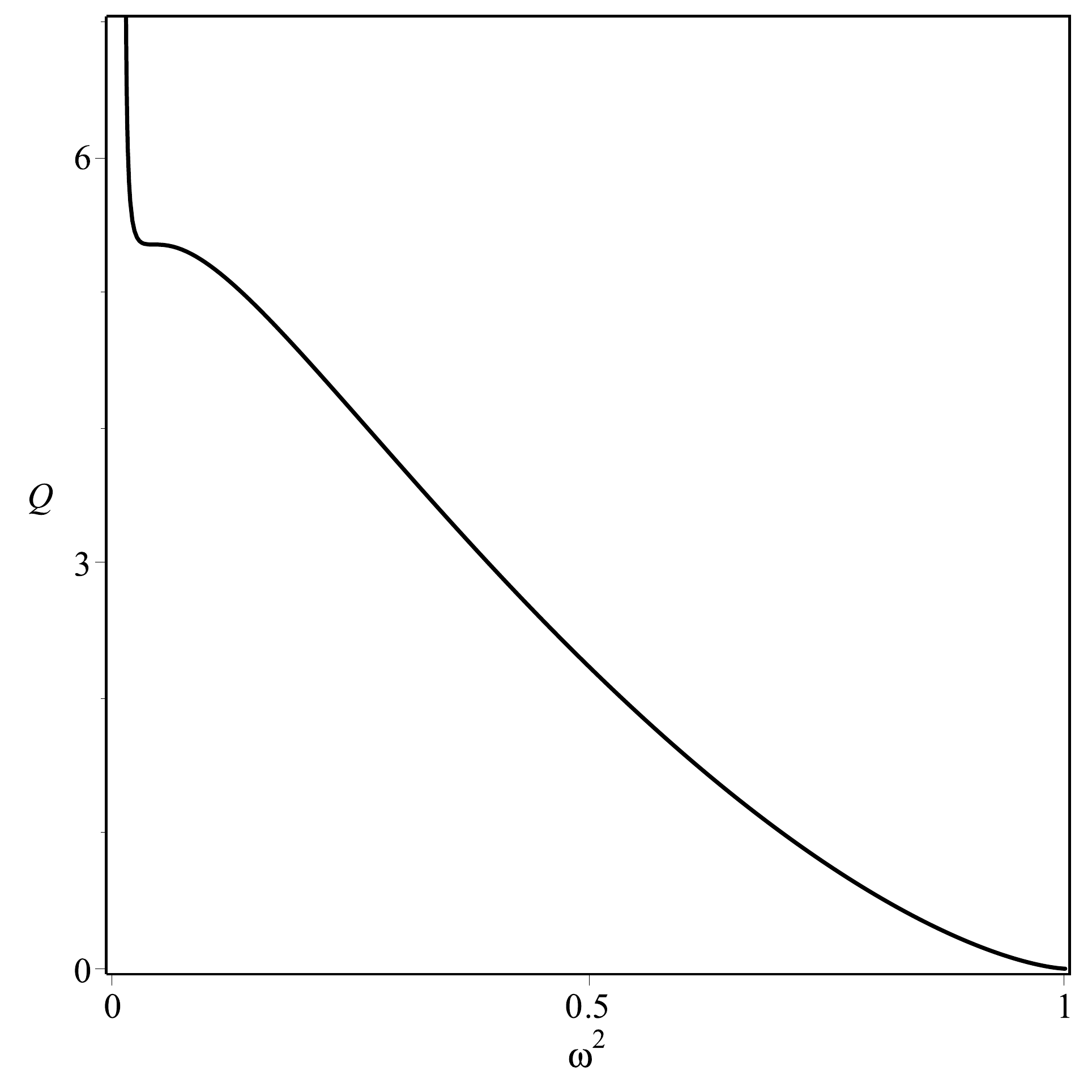}
\caption{The charge as a function of $\omega^2$, for the case $n=1$, with the parameter $a$ as $a_0$ (top panel), $a_1$ (center panel) and $a_2$ (bottom panel).}
\label{fig5}
\end{figure}

To study stability we depict in Fig.~\ref{fig4} and \ref{fig5} the ratio $E/Q$ and $Q$ as a function of $\omega^2$, respectively,
for three distinct values of $a$, as we now explain. We start with the lowest value for the parameter $a$, that is, $a= a_0=2/9\approx 0.2222222$. For $0=\omega_-<\omega<\tilde{\omega}\approx 0.0752748$ one can see that $E/Q>\omega_+$, which is out of the interval in which the Q-ball is stable. Furthermore, the charge is not monotonically decreasing with $\omega$. Thus, the case $a=a_0$ is unstable classically and quantum mechanically. We continue the investigation taking values of $a$ higher than $a_0$. The ratio $E/Q$ has its peak above $\omega_+$ but, for $a$ increasing, it goes down until we get to $a_1=2/9+0.0004596\approx 0.2226819$, where the peak in $E/Q$ is approximately equal to $\omega_+$. The interesting fact in this case is that the ratio $E/Q$ now is in the allowed range that ensures quantum mechanical stability. Nevertheless, the model is yet classically unstable, because the charge is not monotonically decreasing with $\omega$. We go further on, increasing $a$, and the peak of $E/Q$ in the small $\omega$ region goes down and down, until the concavity of the curve changes at $a_2=2/9+0.0031751\approx 0.2253973$. As we increase $a$ up to $a_2$ the charge tends to infinity for $\omega\to\omega_-$ and has local minimum in the small $\omega$ region, which becomes an inflection point when $a= a_2$. Therefore, for $a > a_2$, the solution is stable, both classically and quantum mechanically. 

As we see, in Fig.~\ref{fig4} we depict the ratio $E/Q$ for the three values of $a$, $a=a_0$, $a=a_1$ and $a=a_2$. We note that quantum mechanical instability appears only in the top panel, because $E/Q$ may overcome $\omega_+$. Moreover, classical stability only appears in the bottom panel of the figure. We also depict in Fig.~\ref{fig5} the charge as a function of $\omega^2$, for the same three values of $a$ $(a_0, a_1, a_2)$, and there we see that the charge only becomes a monotonically decreasing function of $\omega^2$ for $a> a_2$. 

\begin{figure}[t!]
\includegraphics[width=5.8cm]{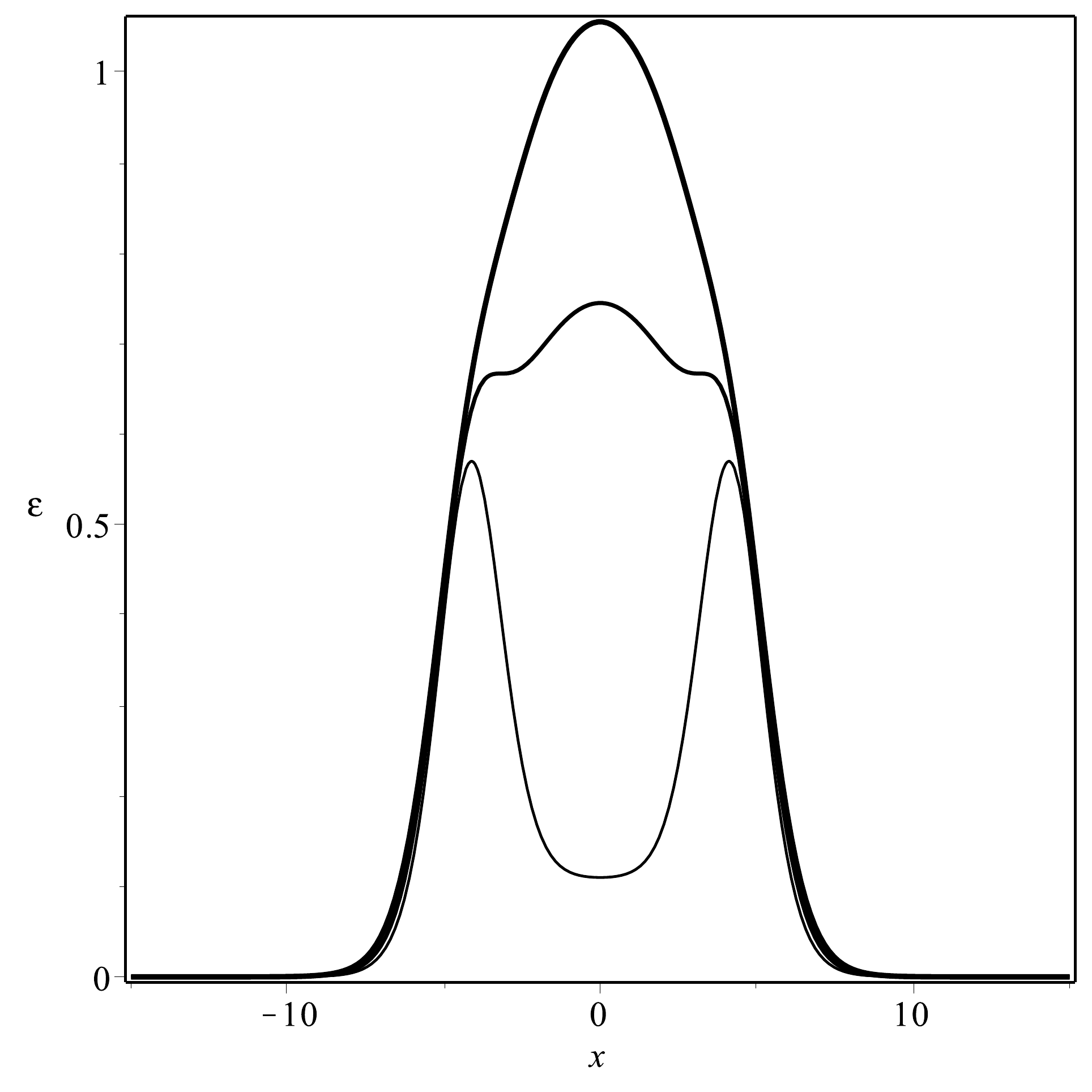}
\caption{The energy density for $a = 9/40, 1/4$ and $11/40$. In each plot, we take $\omega^2 = \omega_-^2 + 10^{-3}$, and the thickness of the line increases as $a$ increases.}
\label{splitting}
\end{figure}

We have done a closer inspection on the energy density, searching for any possible change of behavior. Setting $n=1$ in Eq.~\eqref{ac}, we see that the energy density tends to split when $a$ is in the interval $a_2< a< a_3 = 1/4$, with $\omega$ in the range of Eq.~\eqref{omegac}. We start with $a=a_2$, and as we increase $a$, the central well in the energy density becomes a hill, making the splitting to vanish. We illustrate this in Fig.~\ref{splitting}, where we depict the energy density for $a = 9/40, 1/4$ and $11/40$, using $\omega^2 = \omega_-^2 + 10^{-3}$. It is interesting to note that the tendency to split starts to appear for $a\in (a_2,a_3)$, for $\omega$ in the interval \eqref{omegac}, so it is inside the range where the Q-ball is stable, both quantum mechanically and classically.

\subsection{The case $n=2$}
\label{sec:n=2}

\begin{figure}[htb!]
\includegraphics[width=6.0cm]{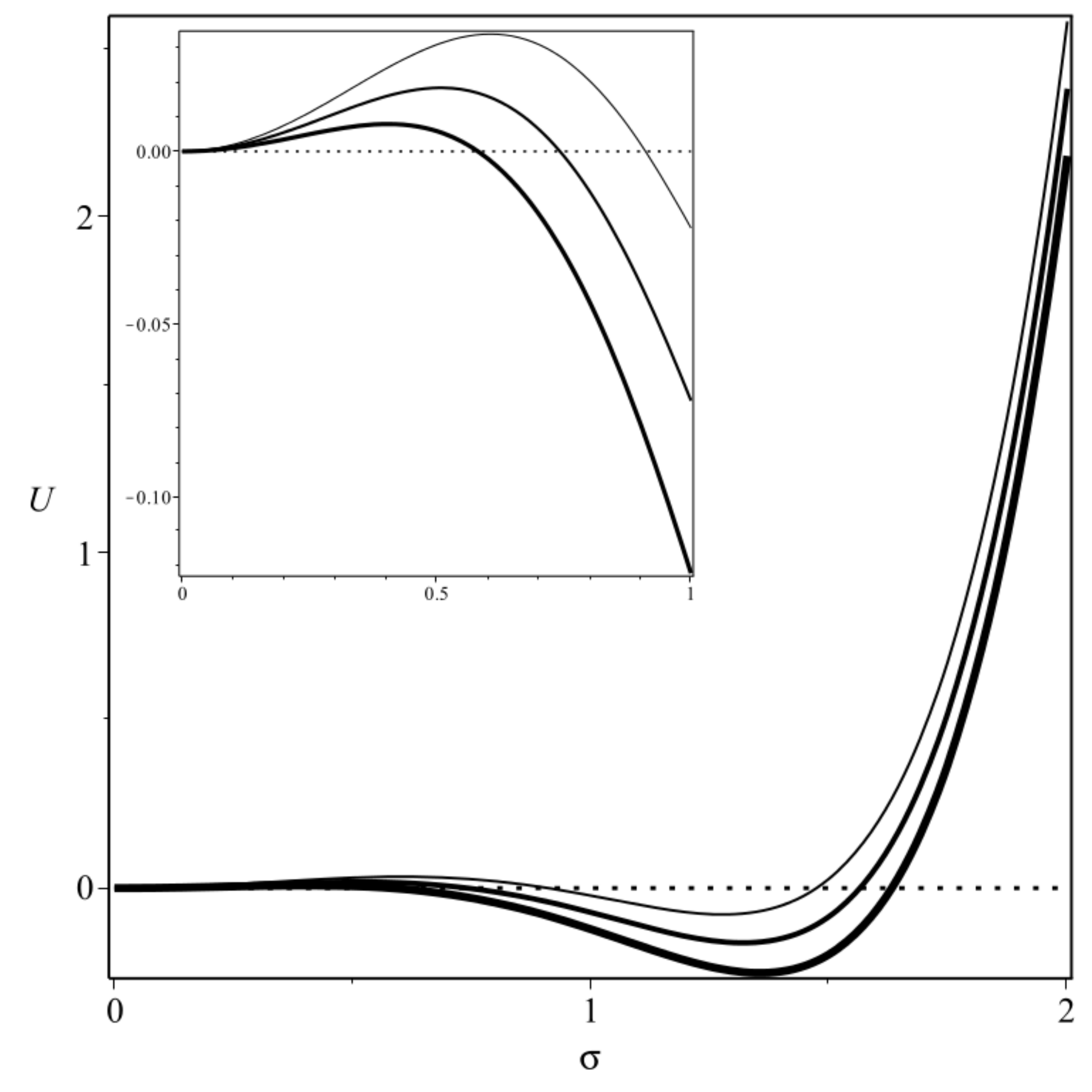}
\caption{The effective potential \eqref{veff2} depicted for $a=4/9$ and $\omega^2=0.6,0.7$ and $0.8$. The thickness of the line increases with $\omega^2$. In the inset, one shows the behavior of the effective potential for $\sigma\in [0,1]$.}\label{fig7}
\end{figure}

We take $n=2$ in Eq.~\eqref{veffn} to get to
\be\label{veff2}
U(\sigma) = \frac12 (1-\omega^2)\sigma^2 -\frac13 \sigma^{4} + \frac14 a\,\sigma^{6}.
\ee
This potential is plotted in Fig.~\ref{fig7}. With the right scaling of parameters, one can show that this is the same case studied in \cite{tdlee}, but here we go further and add new features to the model.

\begin{figure}[htb!]
\includegraphics[width=4.24cm]{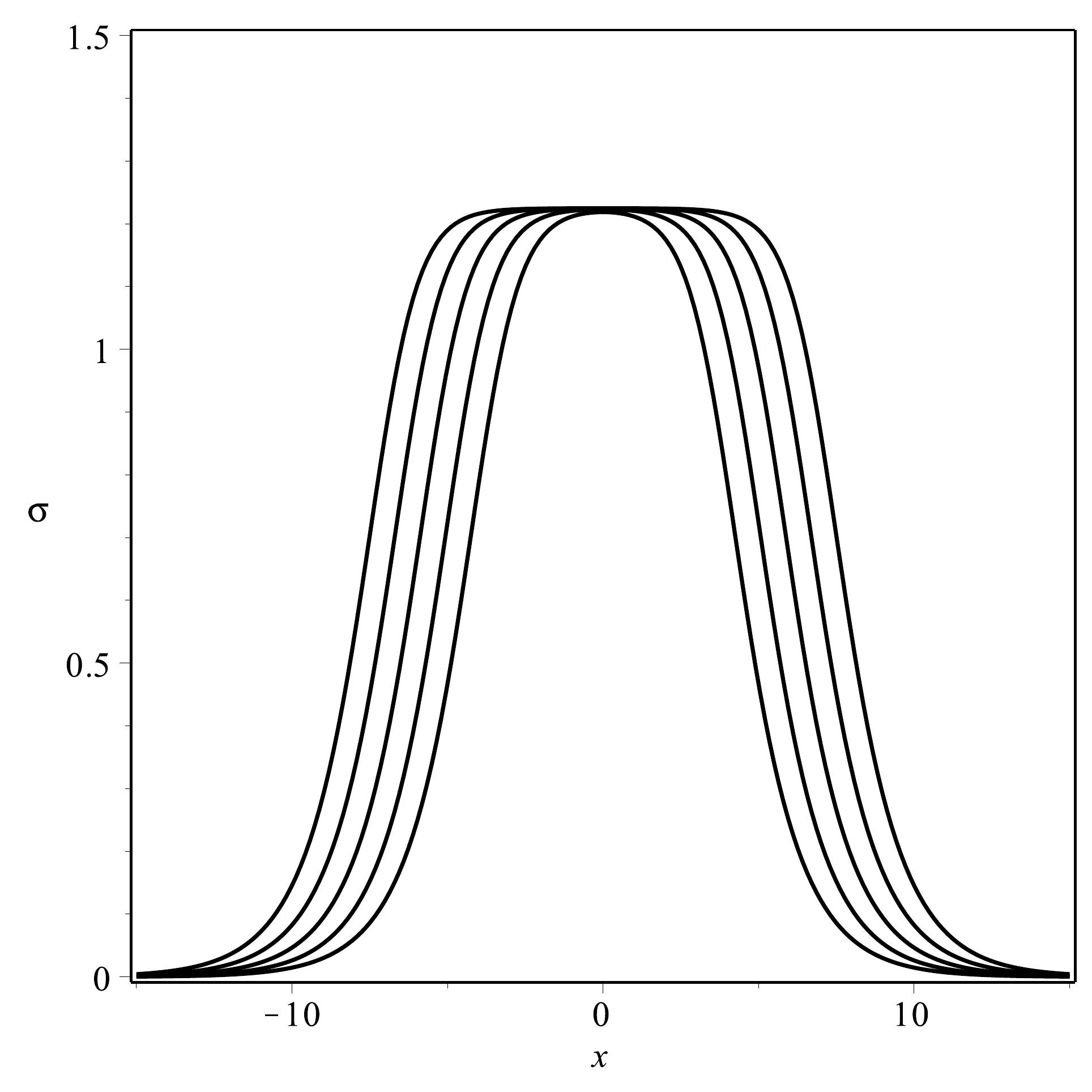}
\includegraphics[width=4.24cm]{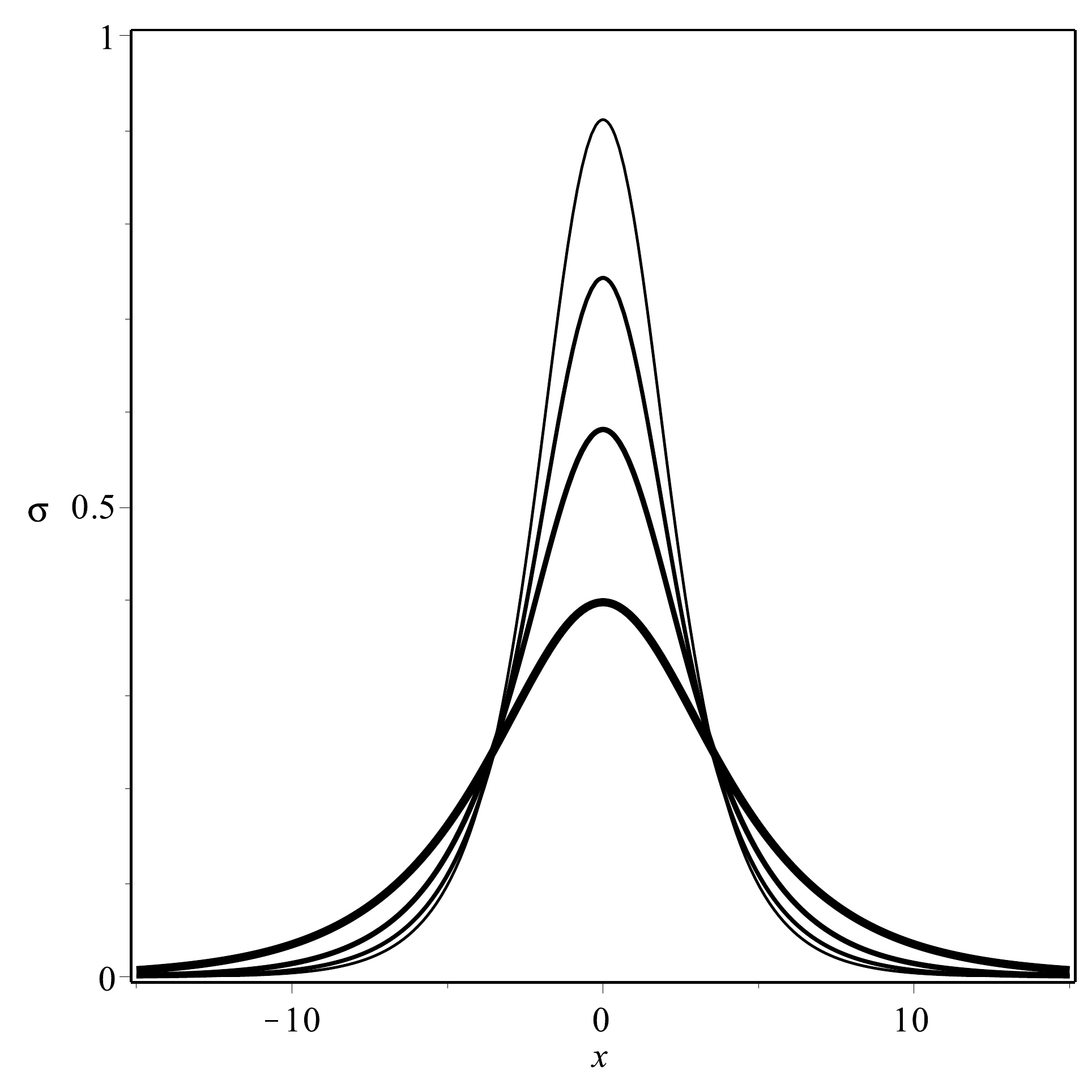}
\caption{The solution \eqref{soln} depicted for $n=2$ and $a=4/9$, with $\omega^2=0.5+5\epsilon$, $\epsilon=10^{-9},10^{-8},10^{-7},10^{-6}$ and $10^{-5}$ (left), and with $\omega^2=0.6, 0.7,0.8$ {\rm and} $0.9$ (right). The plateau in the left panel increases as $\omega$ approaches $\omega_-$. The thickness of the line in the right panel increases as $\omega^2$ increases.}\label{fig8}
\end{figure}

The solution can be found setting $n=2$ in Eq.~\eqref{soln}, which is depicted in Fig.~\ref{fig8} for $a=4/9$ ($\omega^2_-=0.5$), for several values of $\omega$ obeying Eq.~\eqref{condomega}. In Fig.~\ref{fig8}, in the left panel we can see the plateau for $\omega\approx\omega_-$ and in the right panel it is shown that the amplitude of the solution decreases as $\omega$ increases toward $\omega_+$.
 
\begin{figure}[htb!]
\includegraphics[width=5.4cm]{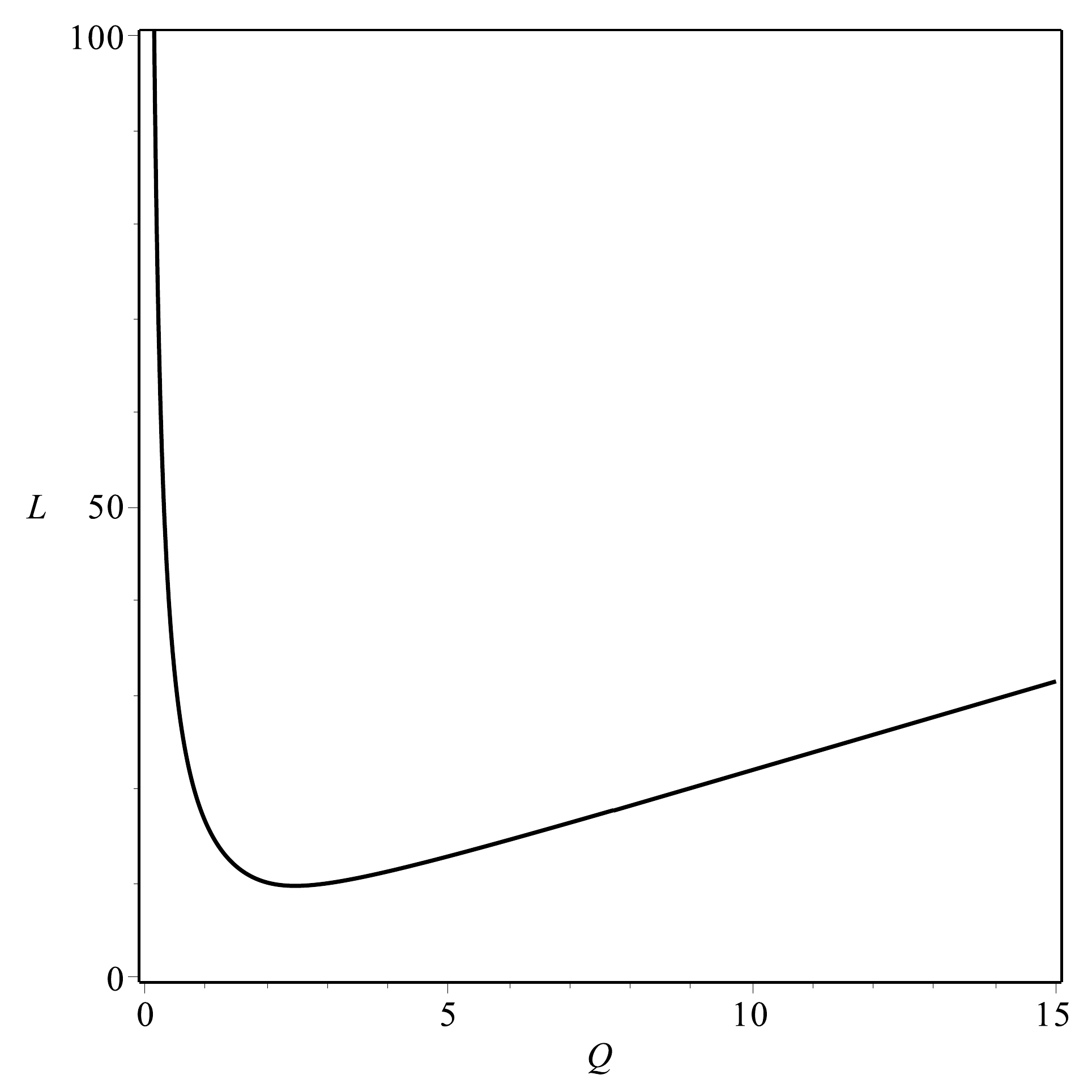}
\caption{The behavior of the width \eqref{widthn} for $n=2$ as a function of the charge \eqref{charge2} for $a=4/9$.}\label{chargewidth2}
\end{figure}

The charge \eqref{chargen} simplifies for $n=2$, becoming
\be\label{charge2}
Q=\sqrt{\frac2a}\,\omega\,\text{arctanh}\left({3\,\sqrt{\frac{(1-\omega^2)\,a}{2}}}\right).
\ee
We see that $Q\to0$ as $\omega\to\omega_+$, for any $a$. Also, specifically for $a=2/9$ we have $\omega_-=0$, which makes $Q\to0$ for
$\omega\to\omega_-=0$. For $a>2/9$ we have $Q\to\infty$ for $\omega\to\omega_-$. The width can be easily obtained from Eq.~\eqref{widthn}. In Fig.~\ref{chargewidth2}, we display the width as a function of the charge for $a=4/9$. The minimum of this curve can be calculated numerically for each value of $a$. In particular, for $a=4/9$, the minimum appears for $\omega_m \approx 0.7655362$, $Q \approx 2.4796264$ and $L\approx 9.6671940$. We define this as the point that separates small Q-balls from large Q-balls.

The kinetic \eqref{ken} and gradient \eqref{gen} energies simplify to
\ben
E_k &=&\frac{\omega^2}{\sqrt{2a}}\,\text{arctanh}\left({3\,\sqrt{\frac{(1-\omega^2)\,a}{2}}}\,\right)\\
E_g &=& \frac14 \sqrt{\frac2a} \frac{e^{8b}-8b\,e^{4b}-1}{\left(e^{4b}-1\right)^2} \left(1-\omega^2\right)
\een

\begin{figure}[htb!]
\includegraphics[width=5.4cm]{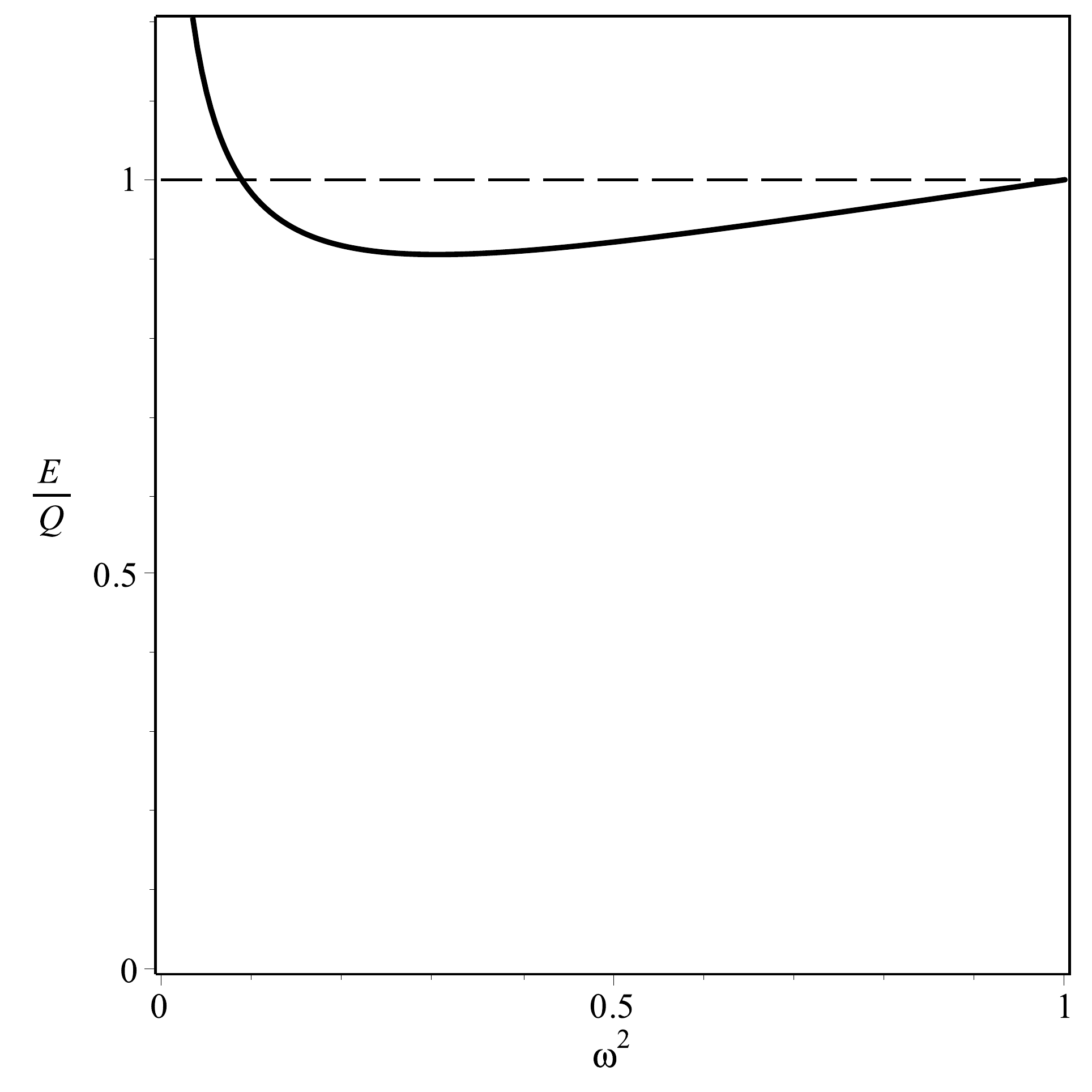}\\
\includegraphics[width=5.4cm]{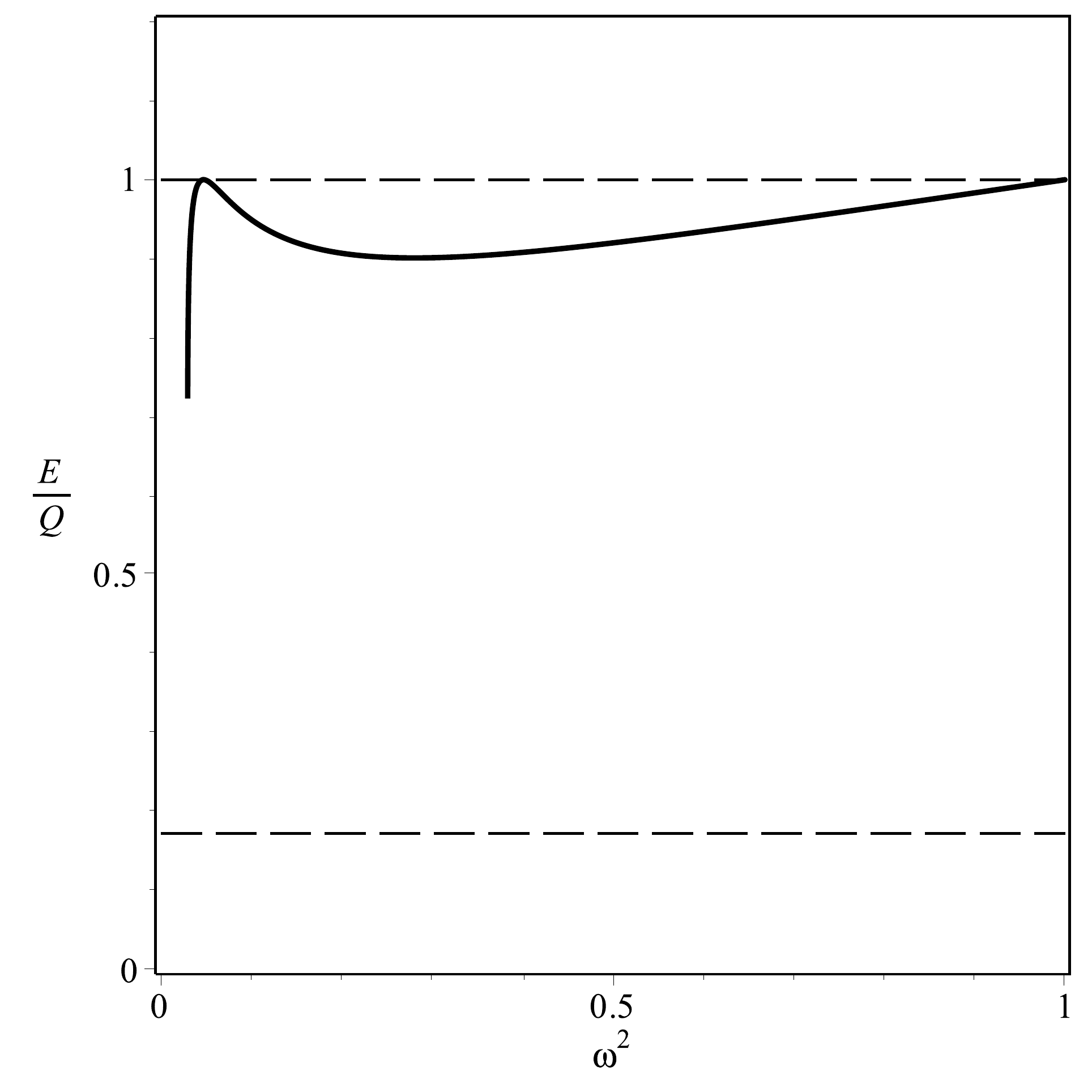}\\
\includegraphics[width=5.4cm]{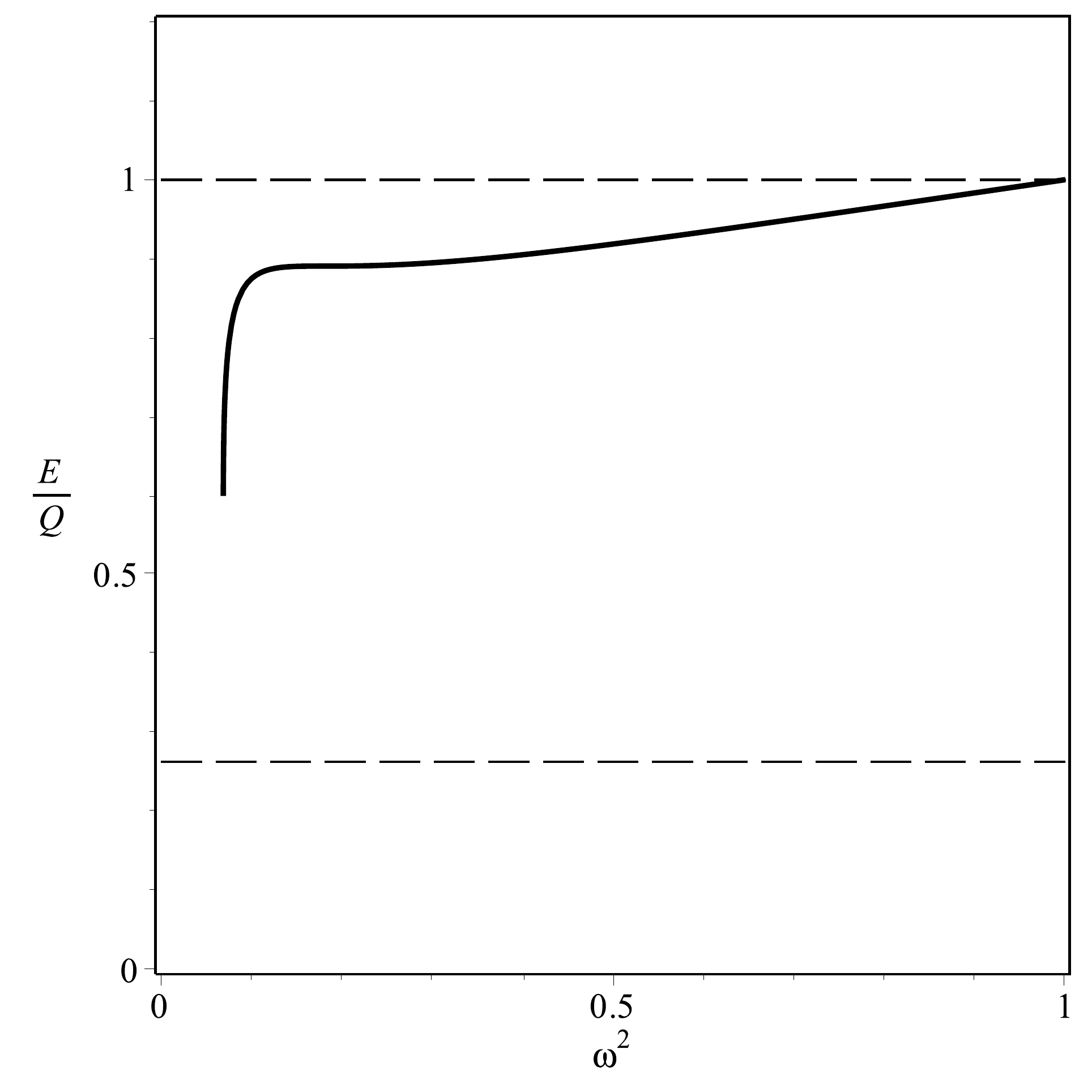}
\caption{The ratio $E/Q$ as a function of $\omega^2$, for the case $n=2$, with the parameter $a$ as $a_0$ (top panel), $a_1$
(center panel) and $a_2$ (bottom panel). The region in between the two dashed horizontal lines assures quantum mechanical stability of the Q-Ball.}\label{fig10}
\end{figure}

\begin{figure}[t!]
\includegraphics[width=5.2cm]{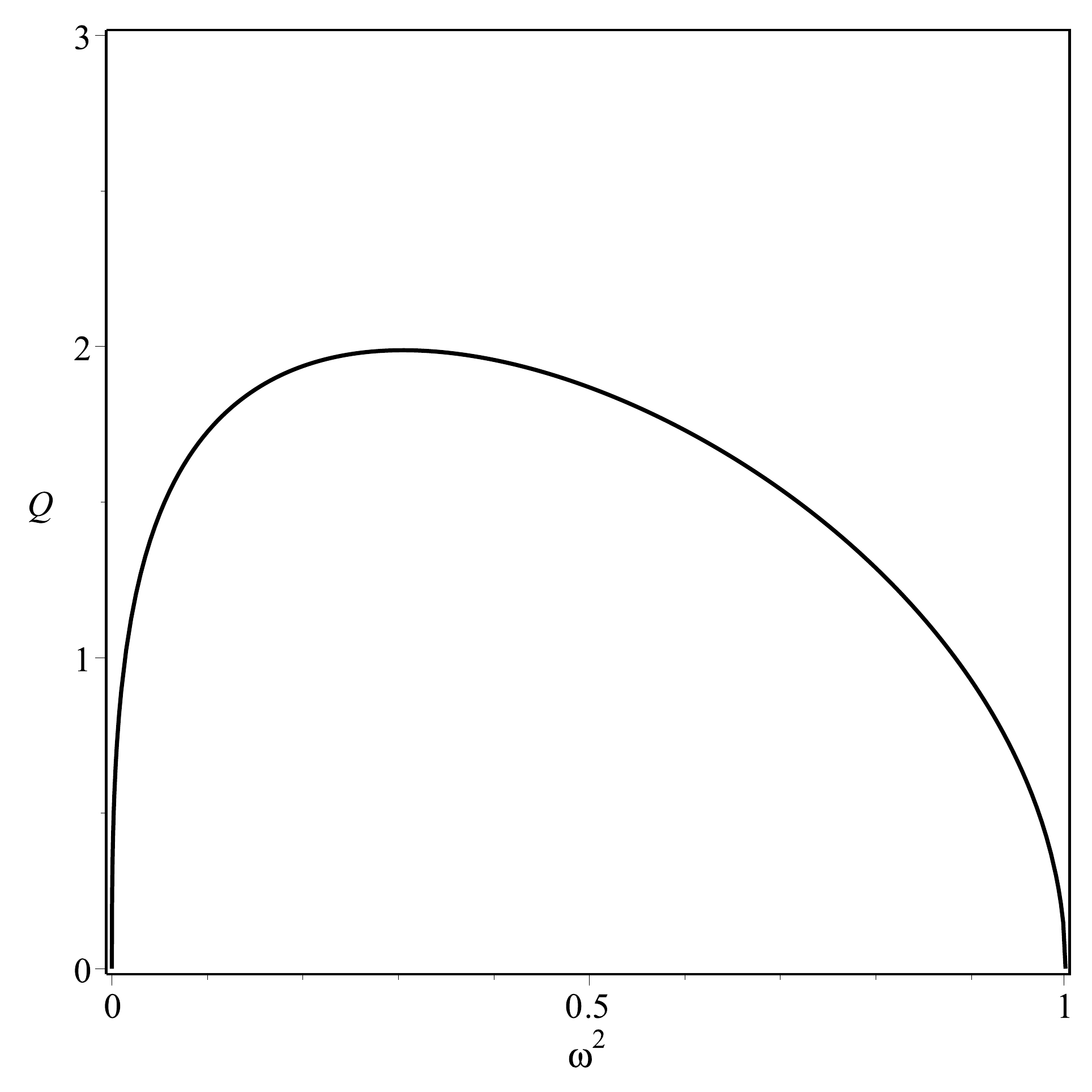}
\includegraphics[width=5.2cm]{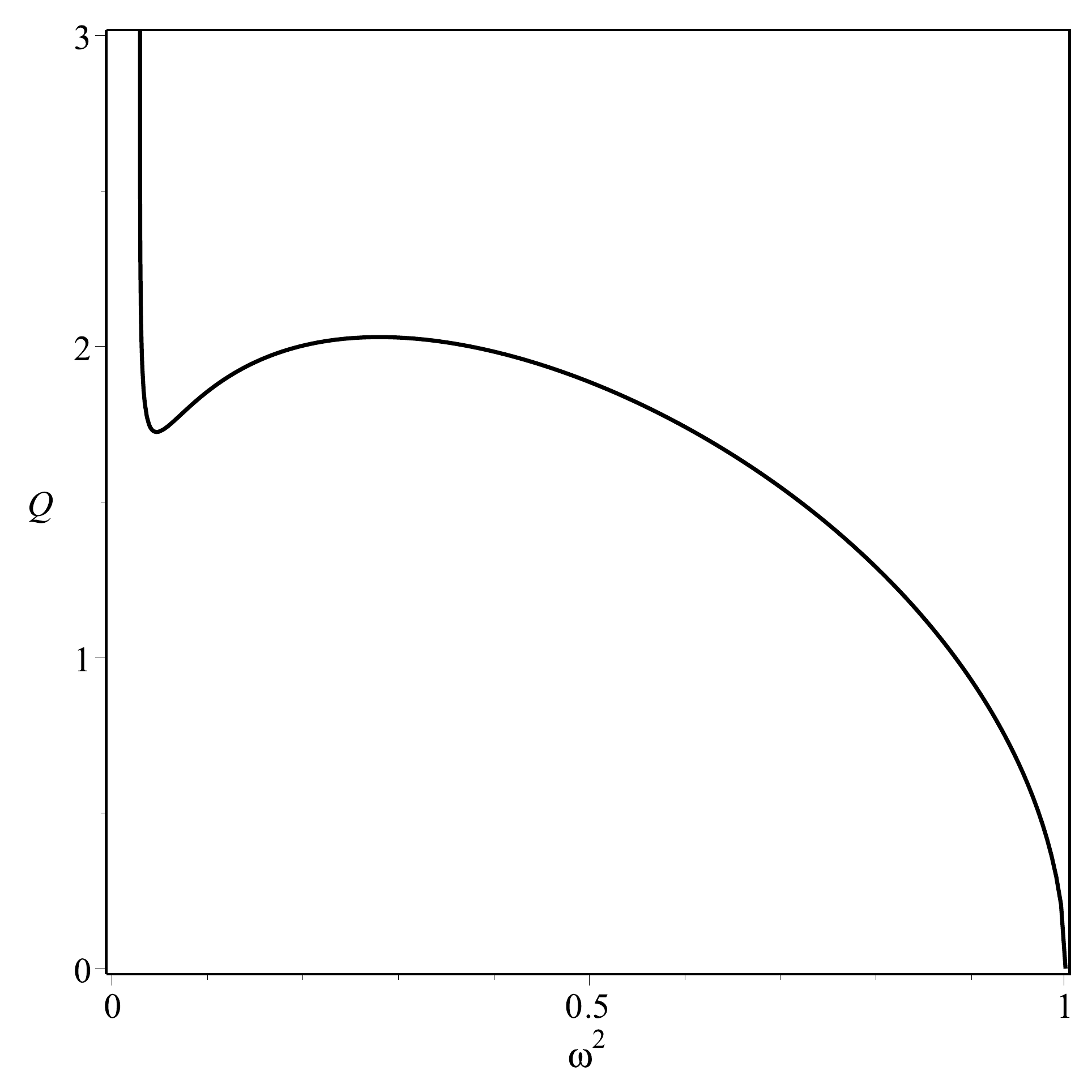}
\includegraphics[width=5.2cm]{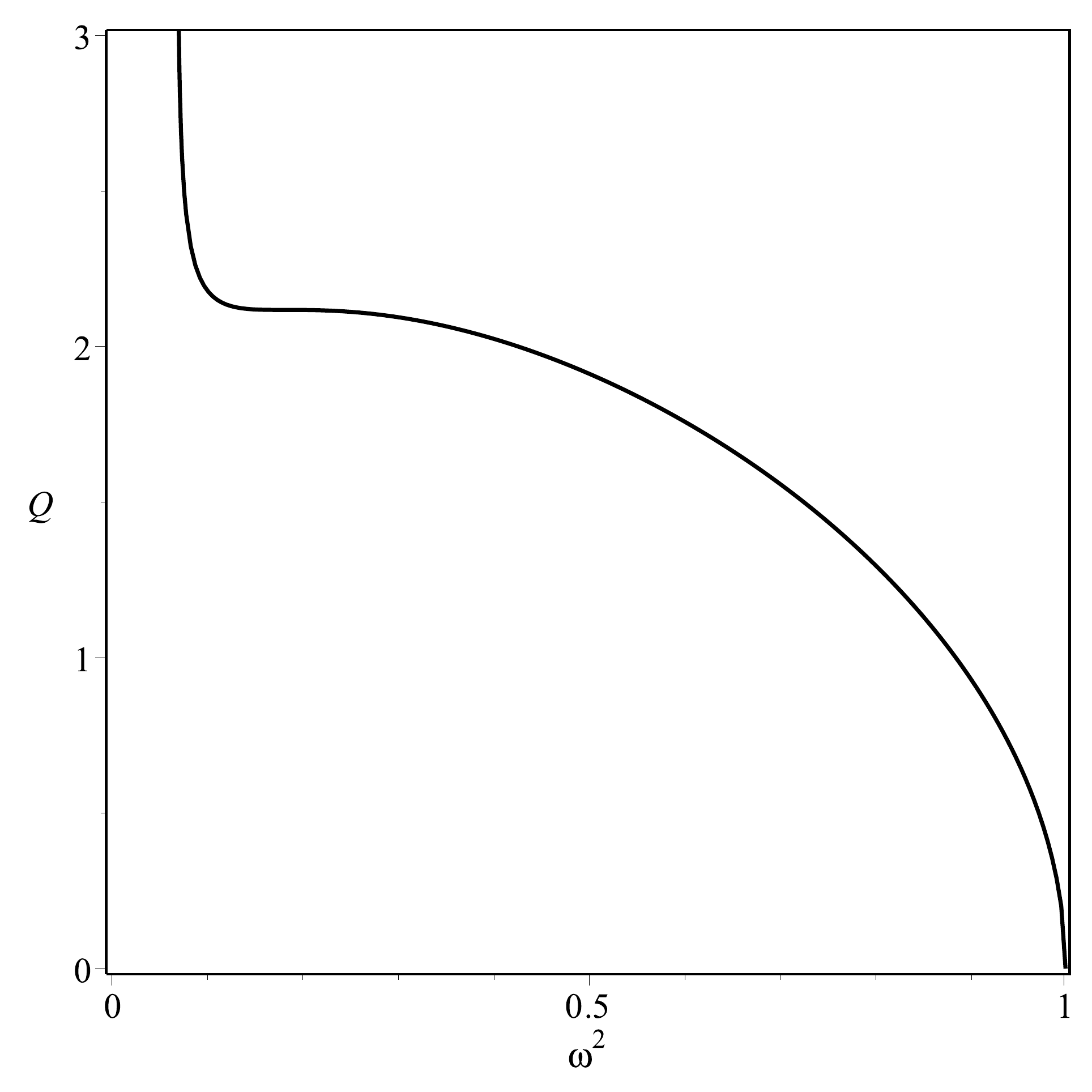}
\caption{The charge as a function of $\omega^2$, for the parameter $a$ as $a_0$ (top panel), $a_1$ (center panel) and $a_2$ (bottom panel).}
\label{fig11}
\end{figure}

To study stability as before. We depict in Fig.~\ref{fig10} and \ref{fig11} the ratio $E/Q$ and $Q$ as a function of $\omega^2$, respectively, for three distinct values of $a$, as we now explain. We start with $a= a_0=2/9\approx 0.2222222$. For $0=\omega_-<\omega<\tilde{\omega}\approx 0.2985278$ one can see that $E/Q>\omega_+$, which is out of the interval in which the Q-ball is stable. Also, the charge is not monotonically decreasing with $\omega$, and so the case $a=a_0$ is unstable classically and quantum mechanically. We continue the investigation, and for $a$ higher than $a_0$, the ratio $E/Q$ has its peak above $\omega_+$ but, for $a$ increasing, it goes down until we get to $a_1=2/9+0.0067204\approx 0.2289426$, where the peak in $E/Q$ is approximately equal to $\omega_+$. The interesting fact in this case is that the ratio $E/Q$ now is in the allowed range that ensures quantum mechanical stability. Nevertheless, the model is yet classically unstable, because the charge is not monotonically decreasing with $\omega$. We go further on, increasing $a$, and the peak of $E/Q$ in the small
$\omega$ region goes down and down, until the concavity of the curve changes at $a_2=2/9+0.0164064\approx 0.2386286$. As we increase $a$ up to $a_2$ the charge tends to infinity for $\omega\to\omega_-$ and has local minimum in the small $\omega$ region, which becomes an inflection point when $a=a_2$. Thus, for $a > a_2$, the solution is stable, both classically and quantum mechanically. 

As we see, in Fig.~\ref{fig10} we depict the ratio $E/Q$ for the three values of $a$, $a=a_0$, $a=a_1$ and $a=a_2$. We note that quantum mechanical instability appears only in the top panel, because $E/Q$ may overcome $\omega_+$. Also, classical stability only appears in the bottom panel of the figure. We also depict in Fig.~\ref{fig11} the charge as a function of $\omega^2$, for the same three values of $a$ $(a_0, a_1, a_2)$, and there we see that the charge only becomes a monotonically decreasing function of $\omega^2$ for $a>a_2$. 

\begin{figure}[t!]
\includegraphics[width=5.8cm]{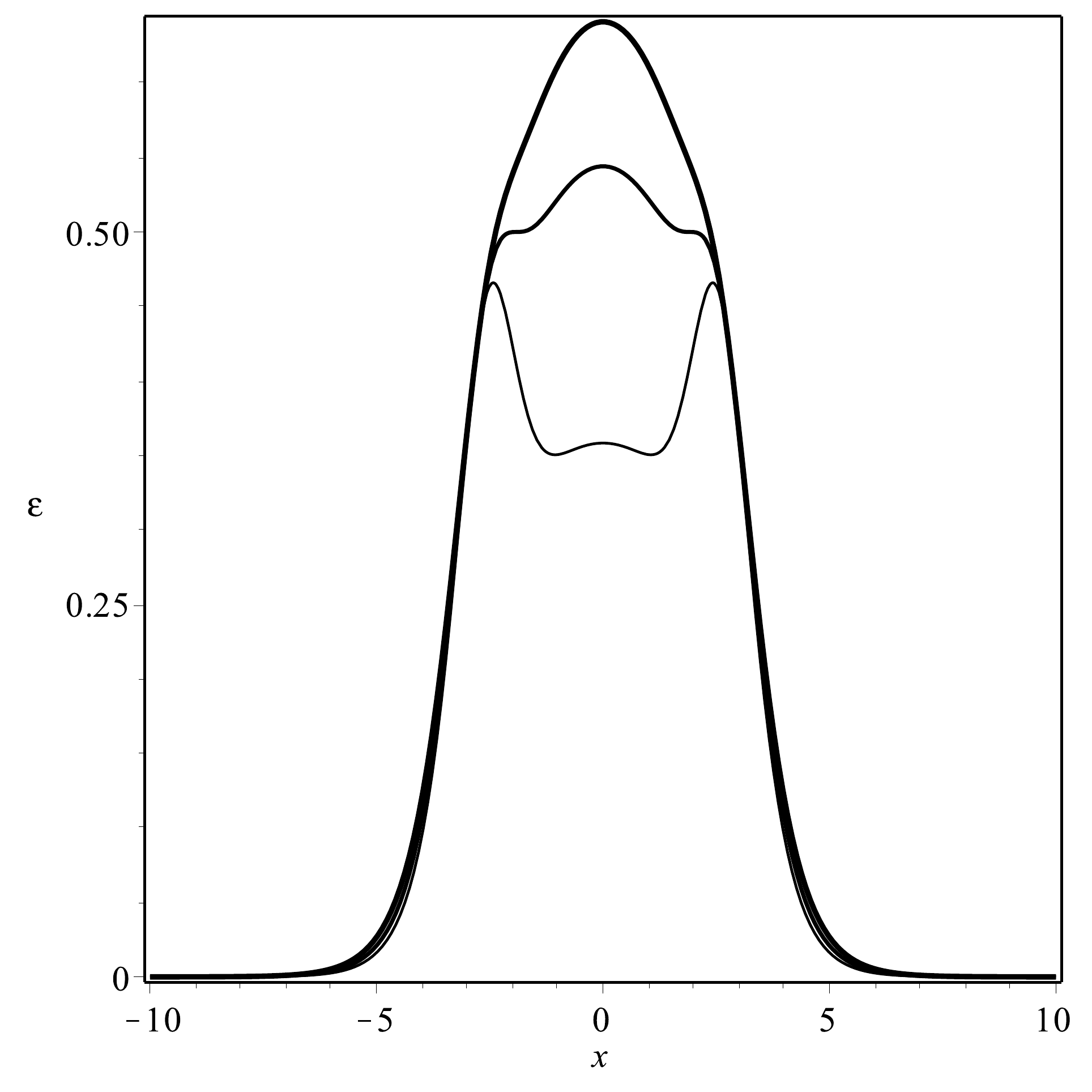}
\caption{The energy density for $a = 7/27, 8/27$ and $1/3$. In each plot, we take $\omega^2 = \omega_-^2 + 10^{-3}$, and the thickness of the line increases as $a$ increases.}
\label{splitting2}
\end{figure}

As in the previous model, we have studied the behavior of the energy density. Setting $n=2$ in Eq.~\eqref{ac}, we see that the energy density tends to split when $a$ is in the interval $a_2< a< a_3 = 8/27$, with omega in the range of Eq.~\eqref{omegac}. We start with $a=a_2$, and as $a$ increases, the central well that appears in the energy density becomes a hill, making the splitting to vanish. We illustrate this in Fig.~\ref{splitting2}, where we depict the energy density for $a = 7/27, 8/27$ and $1/3$, using $\omega^2 = \omega_-^2 + 10^{-3}$. We note that the tendency to split starts to appear for $a\in (a_2,a_3)$, so it is inside the range where the Q-ball is stable, both quantum mechanically and classically.

\subsection{The case of $n=3$}
\label{sec:n=3}

\begin{figure}[htb!]
\includegraphics[width=6.0cm]{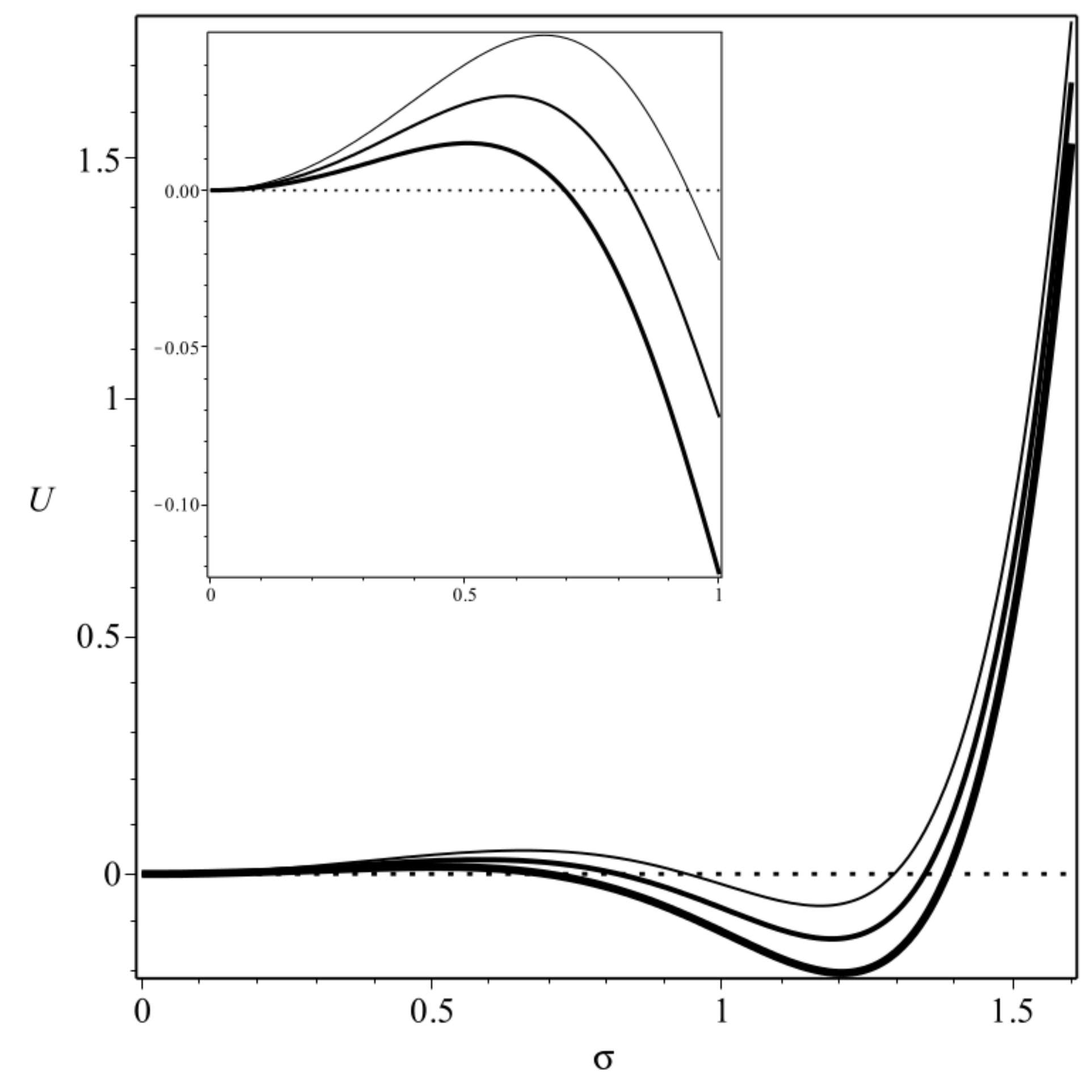}
\caption{The effective potential \eqref{veff3} depicted for $a=4/9$ and $\omega^2=0.6,0.7$ and $0.8$. The thickness of the line increases with $\omega^2$. In the inset, one shows the behavior of the effective potential for $\sigma\in [0,1]$.}\label{fig13}
\end{figure}

We take $n=3$ in Eq.~\eqref{veffn} to get the potential
\be\label{veff3}
U(\sigma) = \frac12 (1-\omega^2)\sigma^2 -\frac13 \sigma^{5} + \frac14 a\,\sigma^{8}.
\ee
This potential is plotted in Fig.~\ref{fig13}.

\begin{figure}[htb!]
\includegraphics[width=4.20cm]{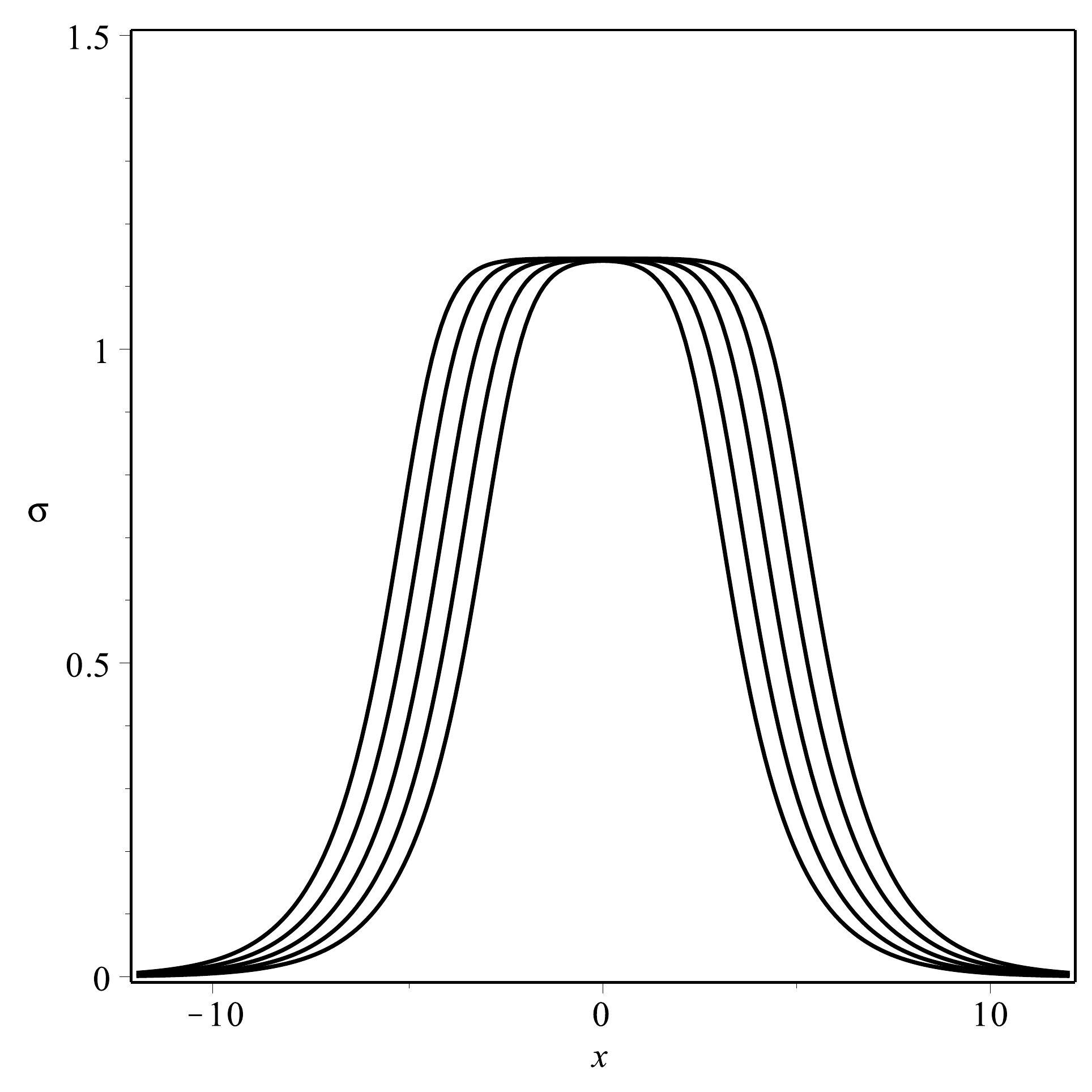}
\includegraphics[width=4.24cm]{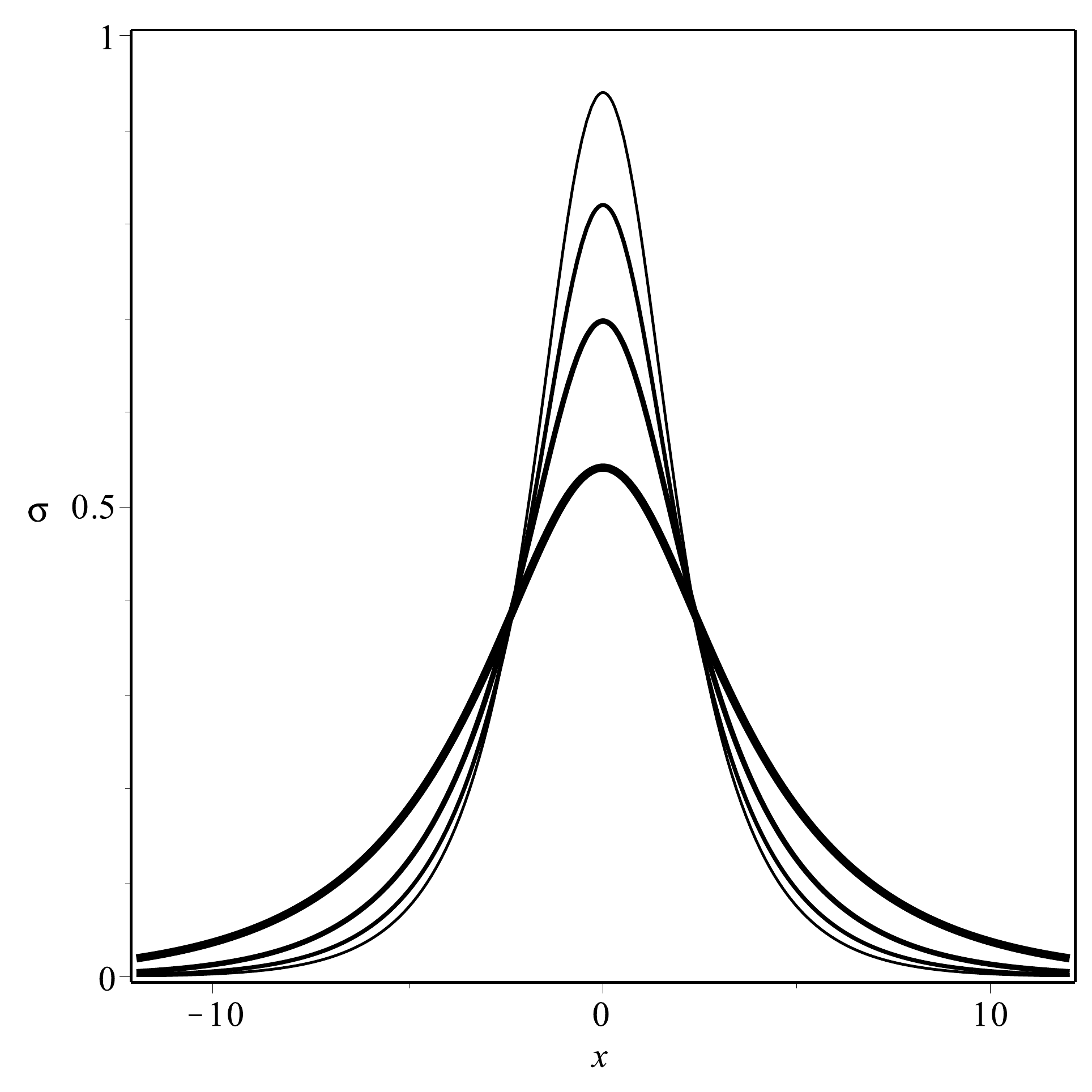}
\caption{The solution \eqref{soln} depicted for $n=3$ and $a=4/9$, with $\omega^2=0.5+5\epsilon$, $\epsilon=10^{-9},10^{-8},10^{-7},10^{-6}$ and $10^{-5}$ (left), and with $\omega^2=0.6, 0.7,0.8$ {\rm and} $0.9$ (right). The plateau in the left panel increases as $\omega$ approaches $\omega_-$. The thickness of the line in the right panel increases as $\omega^2$ increases.}\label{fig14}
\end{figure}

The solution can be found setting $n=3$ in Eq.~\eqref{soln}, which is depicted in Fig.~\ref{fig14} for $a=4/9$ ($\omega^2_-=0.5$), for several values of $\omega$ obeying Eq.~\eqref{condomega}. In this Fig.~\ref{fig14}, we see the plateau for $\omega\approx\omega_-$ in the left panel, and it also shows that the amplitude of the solution decreases as $\omega$ increases toward $\omega_+$, in the right panel.
 
\begin{figure}[htb!]
\includegraphics[width=5.4cm]{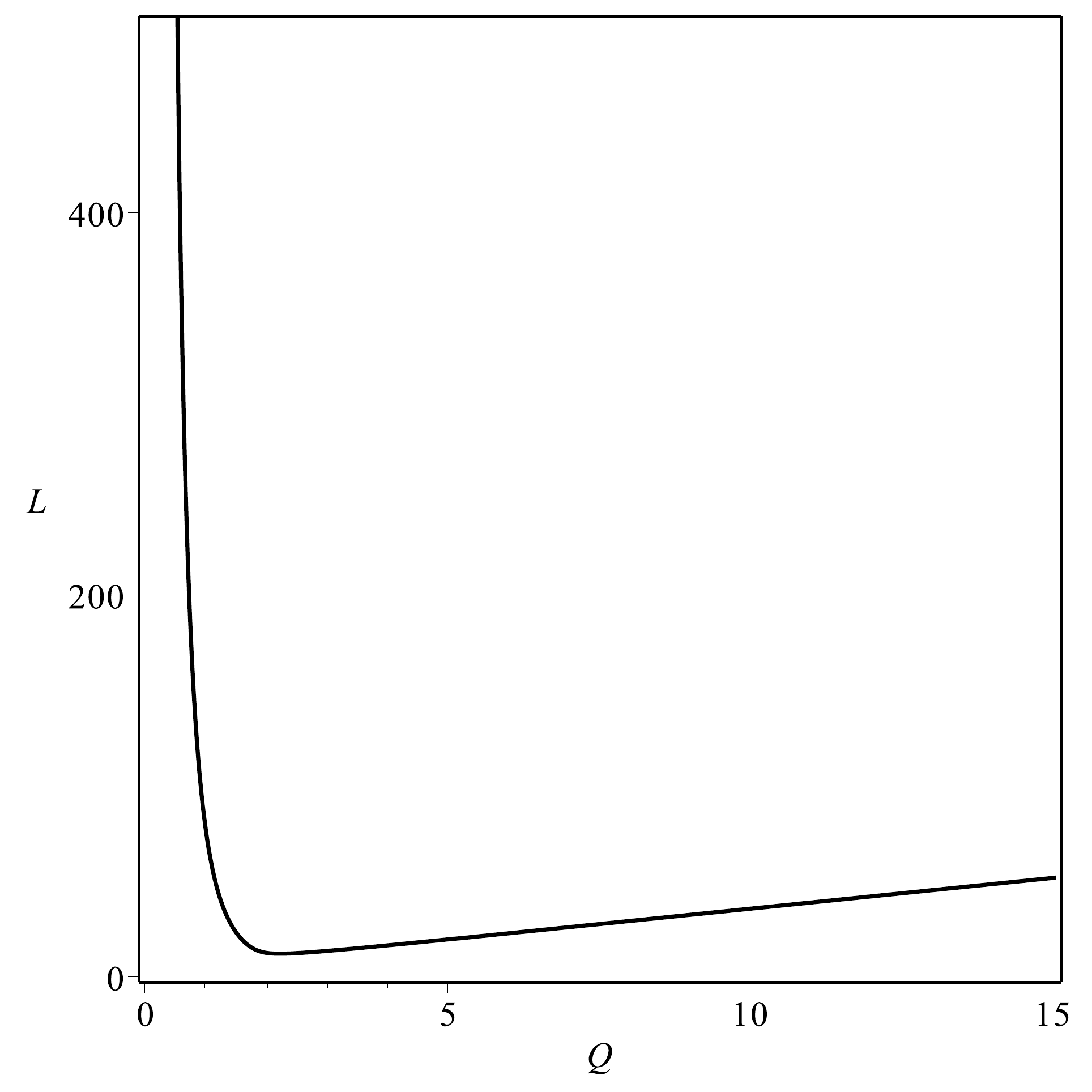}
\caption{The behavior of the width \eqref{widthn} for $n=2$ as a function of the charge \eqref{charge2} for $a=4/9$.}\label{chargewidth3}
\end{figure}

We have not been able to find a simpler expression for \eqref{chargen} in the case $n=3$. However, it is possible to see that $Q\to0$ as $\omega\to\omega_+$, for any $a$. Also, specifically for $a=2/9$ we have $\omega_-=0$, which makes $Q\to0$ for
$\omega\to\omega_-=0$. For $a>2/9$ we have $Q\to\infty$ for $\omega\to\omega_-$. The width can be easily obtained from Eq.~\eqref{widthn}. In Fig.~\ref{chargewidth3}, we display the width as a function of the charge for $a=4/9$. The investigation is similar to the previous one, and for $a=4/9$, the minimum appears for $\omega_m \approx 0.7580576$, $Q \approx 2.2272465$ and $L\approx 12.1294293$. This as the point that separates small Q-balls from large Q-balls in this case.

\begin{figure}[htb!]
\includegraphics[width=5.4cm]{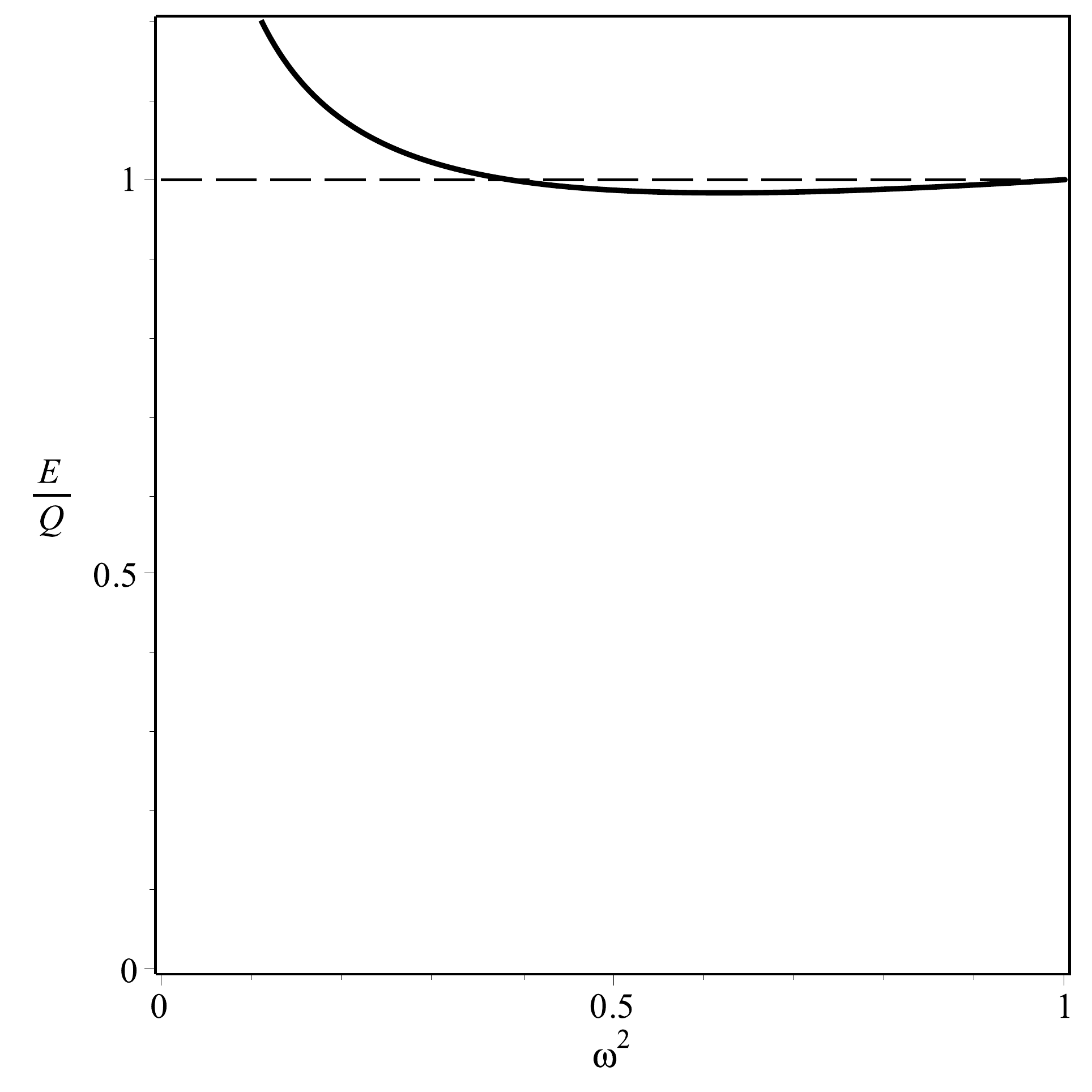}\\
\includegraphics[width=5.4cm]{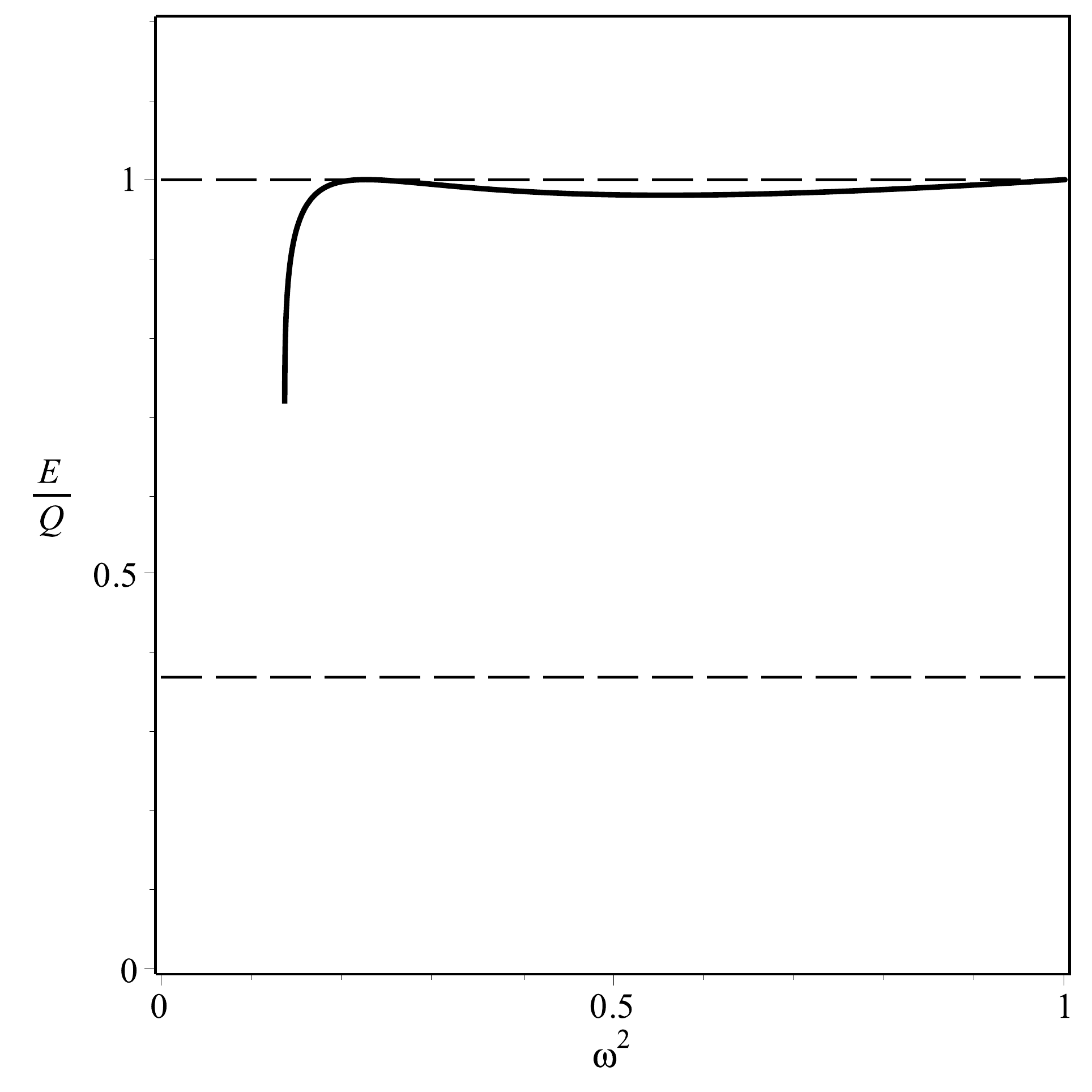}\\
\includegraphics[width=5.4cm]{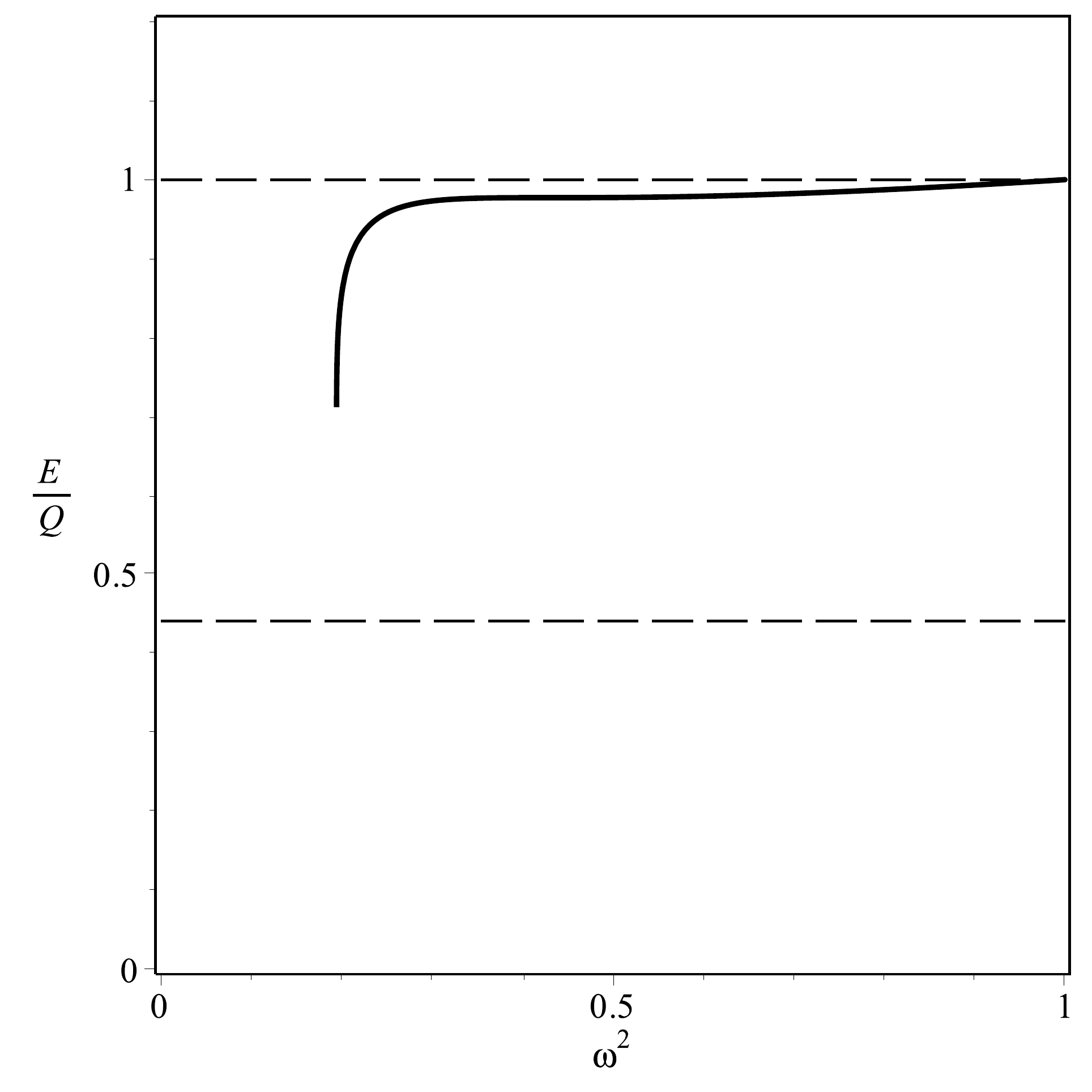}
\caption{The ratio $E/Q$ as a function of $\omega^2$, for the case $n=3$, with the parameter $a$ as $a_0$ (top panel), $a_1$
(center panel) and $a_2$ (bottom panel). The region in between the two dashed horizontal lines assures quantum mechanical stability of the Q-Ball.}\label{fig16}
\end{figure}

\begin{figure}[t!]
\includegraphics[width=5.2cm]{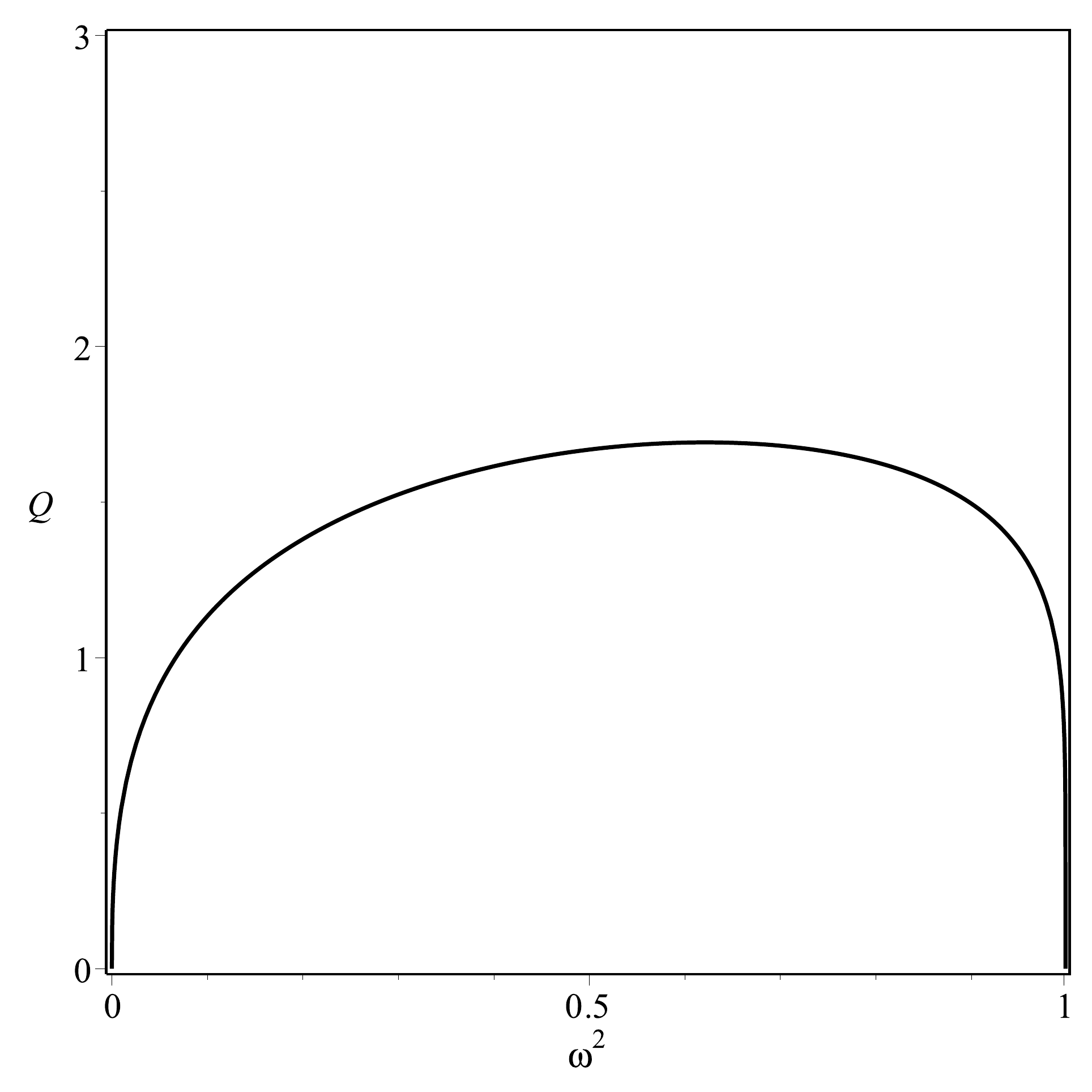}
\includegraphics[width=5.2cm]{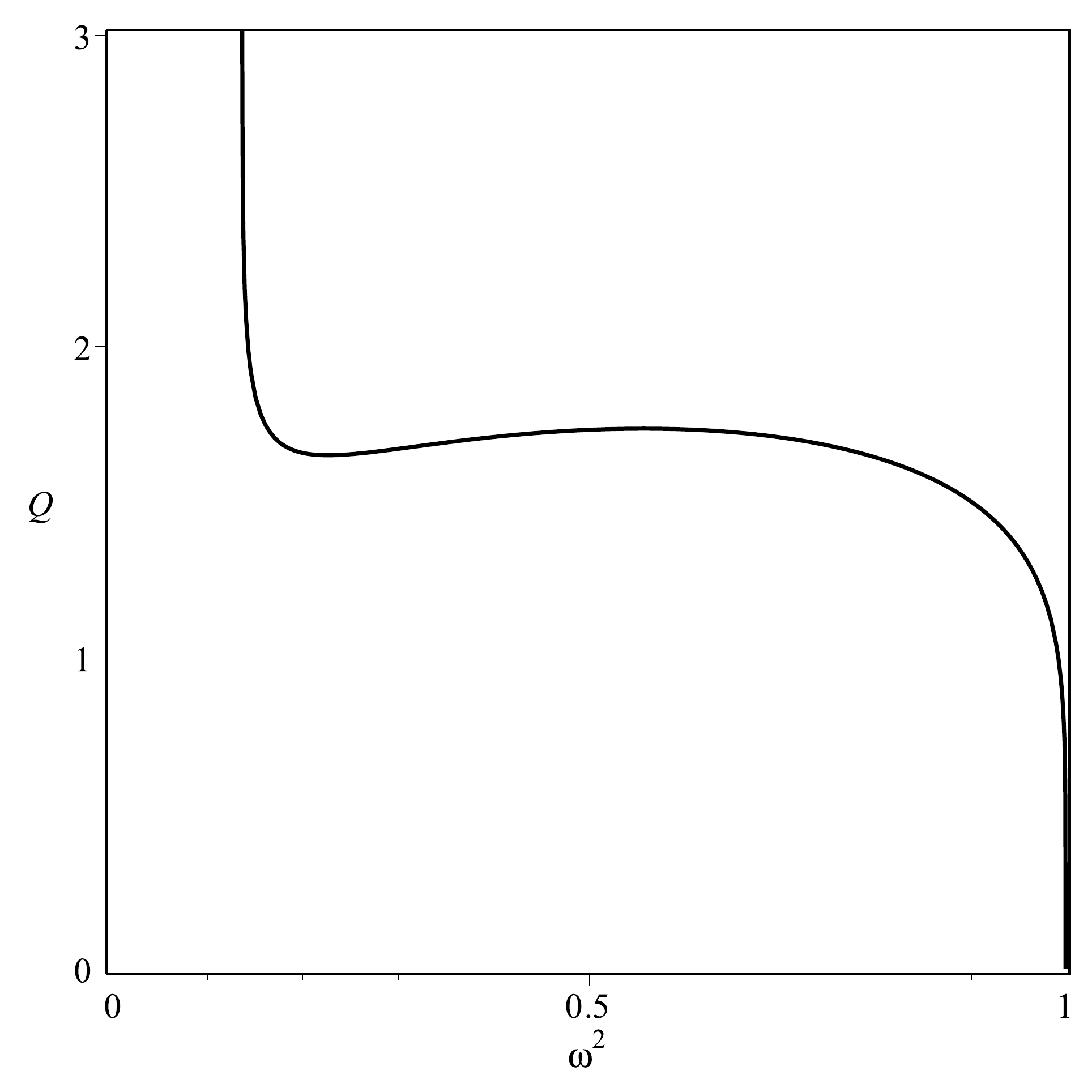}
\includegraphics[width=5.2cm]{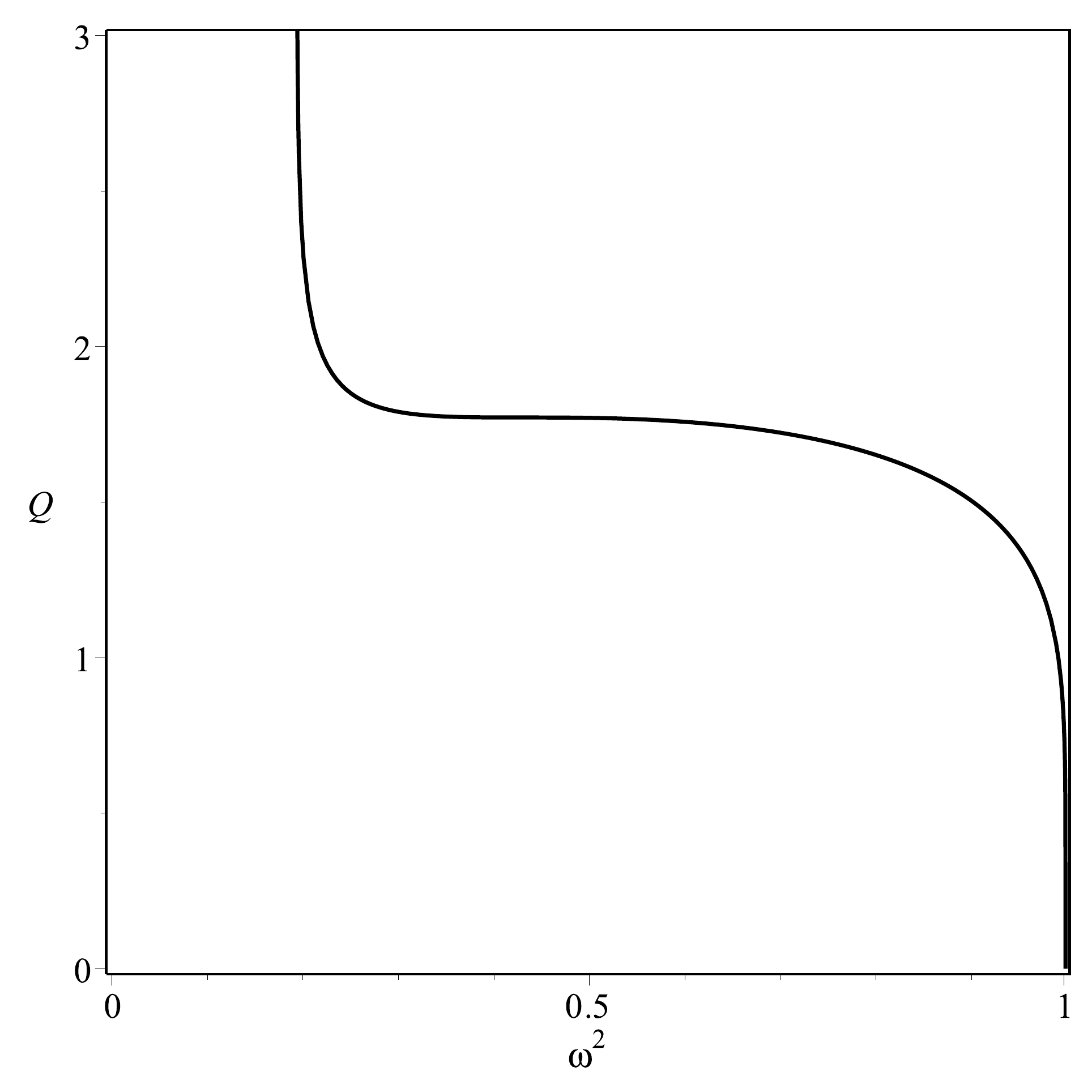}
\caption{The charge as a function of $\omega^2$, for the parameter $a$ as $a_0$ (top panel), $a_1$ (center panel) and $a_2$ (bottom panel).}
\label{fig17}
\end{figure}

To study stability we depict in Fig.~\ref{fig16} and \ref{fig17} the ratio $E/Q$ and $Q$ as a function of $\omega^2$, respectively, for three distinct values of $a$. We start with the lowest value for the parameter $a$, that is, $a= a_0=2/9\approx 0.2222222$. For $0=\omega_-<\omega<\tilde{\omega}\approx 0.6202147$ one can see that $E/Q>\omega_+$, which is out of the interval in which the Q-ball is stable. Also, the charge is not monotonically decreasing with $\omega$, thus the case $a=a_0$ is unstable classically and quantum mechanically. We then consider $a$ higher than $a_0$. The ratio $E/Q$ has its peak above $\omega_+$ but, for $a$ increasing, it goes down until we get to $a_1=2/9+0.0351702\approx 0.2573924$, where the peak in $E/Q$ is approximately equal to $\omega_+$. The ratio $E/Q$ is now in the allowed range to ensure quantum mechanical stability. Nevertheless, the model is yet classically unstable, since the charge is not monotonically decreasing with $\omega$. We keep increasing $a$, and the peak of $E/Q$ in the small $\omega$ region goes down and down, until the concavity of the curve changes at $a_2=2/9+0.0534856\approx 0.2757078$. As we increase $a$ up to $a_2$ the charge goes to infinity for
$\omega\to\omega_-$, and has local minimum in the small $\omega$ region, which becomes an inflection point when $a= a_2$. For $a > a_2$, the solution is then stable, both classically and quantum mechanically. 

In Fig.~\ref{fig10} we depict the ratio $E/Q$ for the three values of $a$, $a=a_0$, $a=a_1$ and $a=a_2$. We note that quantum mechanical instability appears only in the top panel, because $E/Q$ may overcome $\omega_+$. However, classical stability only appears in the bottom panel of the figure. We also depict in Fig.~\ref{fig11} the charge as a function of $\omega^2$, for the same three values of $a$ $(a_0, a_1, a_2)$, and there we see that the charge only becomes a monotonically decreasing function of $\omega^2$ for $a> a_2$. 

\begin{figure}[t!]
\includegraphics[width=5.8cm]{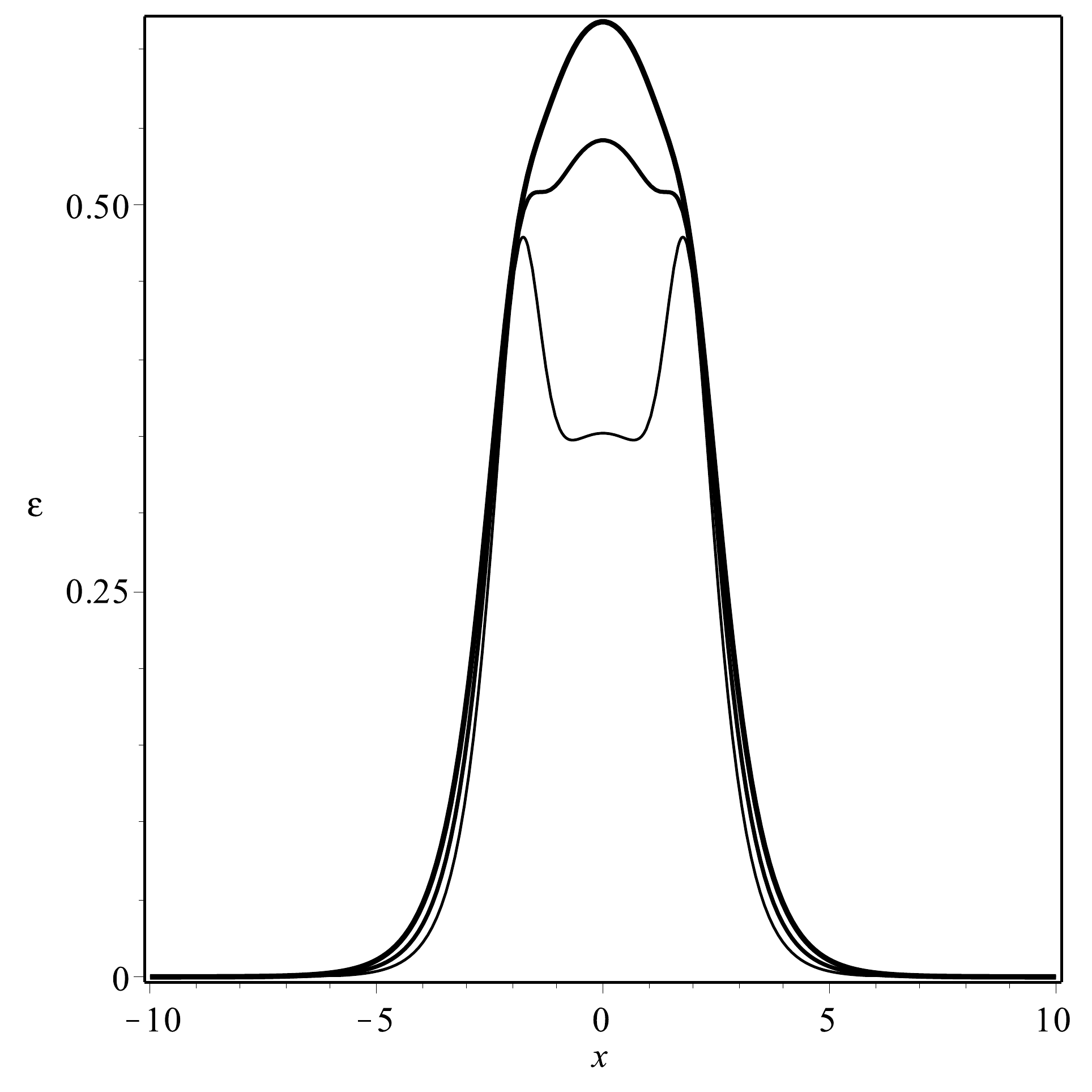}
\caption{The energy density for $a = 5/18, 25/72$ and $5/12$. In each plot, we take $\omega^2 = \omega_-^2 + 10^{-3}$, and the thickness of the line increases as $a$ increases.}
\label{splitting3}
\end{figure}

We have investigated the behavior of the energy density, to see if it splits, as it appeared in the previous models. Taking $n=3$ in Eq.~\eqref{ac}, we see that the energy density tends to split when $a$ is in the interval $a_2< a< a_3 = 25/72$, with omega in the range of Eq.~\eqref{omegac}. We start with $a=a_2$, and as we increase $a$, the well that appears in the central region of the energy density becomes a hill, making the splitting to vanish. We illustrate this in Fig.~\ref{splitting3}, where we depict the energy density for $a = 5/18, 25/72$ and $5/12$, using $\omega^2 = \omega_-^2 + 10^{-3}$. It is interesting to note that the tendency to split starts to appear for $a\in (a_2,a_3)$, so it is inside the range where the Q-ball is stable, both quantum mechanically and classically.

Also, we investigated the case with $n=4$, and we found similar results, so we omit it here.

\section{Ending comments} 
\label{sec:end}

In this work we studied Q-balls in models engendering global U(1) symmetry. We found exact solutions for several distinct values of the parameters that describe the models. In particular, we found that the Q-ball engender the tendency to split, and we also found regions in parameter space where the Q-ball is unstable, quantum mechanically stable, and quantum mechanically and classically stable. The results unveil a behavior which may contribute to enlarge the corresponding dynamics \cite{sut} and modify the scenario for charge-swapping Q-ball interactions \cite{ed}. An interesting issue concerns the tendency to split, which should be further investigated. Another related issue concerns duality, as suggested in \cite{sut1}, where the (Noether) charge of a stationary Q-ball may be dual to the (topological) charge of a static, kinklike structure.

The exact results that we constructed above are of direct importance for Q-balls, and can be used to improve the numerical simulations that appear in the related literature. Perhaps, they may shed light on the present understanding of issues such as baryogenesis, dark matter and other related areas of phenomenology, as we see, for instance, in Refs.~\cite{do,bar} and in references therein.

Another issue of current interest is related to string theory and concerns scenarios involving extra dimensions. A recent work on Q-branes \cite{qbrane} shows that dielectric D-brane systems admit non-abelian Q-ball solutions on their world-volume, and we are now investigating other possible scenarios.

\acknowledgements{We would like to acknowledge the Brazilian agency CNPq for partial financial support. DB thanks support from contracts 455931/2014-3 and 06614/2014-6, MAM thanks support from contract 140735/2015-1, and RM thanks support from contracts 508177/2010-3 and 455619/2014-0.}


\begin{thebibliography}{99}
\bb{b1}N. Manton and P. Sutcliffe, {\it Topological Solitons} (Cambridge University Press, 2004).
\bb{b2}L. Wilets, {\it Nontopological Solitons} (World Scientific, 1989).
\bibitem{tdlee} T.D. Lee and Y. Pang, Phys. Rep. {\bf221}, 251 (1992).
\bibitem{coleman} S. Coleman, Nucl. Phys. B {\bf262}, 263 (1985); erratum {269}, 744 (1986).
\bb{kolb}J.A. Frieman, G.B. Gelmini, M. Gleiser and E.W. Kolb, Phys. Rev. Lett. {\bf60}, 2101 (1988).
\bb{41}J.M. Cerver\'o and P.G. Est\'evez, Phys. Lett. B {\bf176}, 139 (1986).
\bb{41a}C.N. Kumar and A. Khare, J. Phys. A {\bf20}, L1219 (1987).
\bb{42}T.I. Belova and A.E. Kudryavtsev, Sov. Phys. JETP {\bf68}, 7 (1989).
\bibitem{kusenko} A. Kusenko, Phys. Lett. B {\bf404}, 285 (1997).
\bb{prl}A. Kusenko, V. Kuzmin, M. Shaposhnikov, and P.G. Tinyakov, Phys. Rev. Lett. {\bf80}, 3185 (1998). 
\bb{dm}A. Kusenko and M. Shaposhnikov, Phys Lett. B {\bf417}, 99 (1998).
\bb{q1}K. Enqvist and J. McDonald, Phys. Lett. B {\bf425}, 309 (1998); Nucl. Phys. B {\bf538}, 321 (1999).
\bibitem{tuomas} T. Multam\"aki and I. Vilja, Nucl. Phys. B {\bf574}, 130 (2000).
\bibitem{minos} M. Axenides, S. Komineas, L. Perivolaropoulos and M. Floratos, Phys. Rev. D {\bf61}, 085006 (2000).
\bb{q2}S. Kasuya and M. Kawasaki, Phys. Rev. D {\bf62}, 023512 (2000);
Phys. Rev. D {62}, 023512 (2000); Phys. Rev. D {\bf64}, 123515 (2001).
\bb{sut}R.A. Battye and P.M. Sutcliffe, Nucl. Phys. B {\bf590}, 329 (2000).
\bb{ku}A. Kusenko and P.J. Steinhardt, Phys. Rev. Lett. {\bf87}, 141301 (2001).
\bb{sut1}P. Bowcock, D. Foster, and P. Sutcliffe, J. Phys. A {\bf42}, 085403 (2009).
\bb{ed}E.J. Copeland, P.M. Saffin, and  S.-Y.  Zhou, Phys. Rev. Lett. {\bf113}, 231603 (2014).
\bb{sta1}A. Tranberg and D.J. Weir, JHEP {\bf1404}, 184 (2014).
\bb{sta2}E.Ya. Nugaev and M.N. Smolyakov, JHEP {\bf1407}, 009 (2014). 
\bb{ad}I. Affleck and M. Dine, Nucl. Phys. B {\bf249}, 361 (1985).
\bb{rh}K. Enqvist, S. Kasuya, and A. Mazumdar, Phys. Rev. Lett. {\bf89}, 091301 (2002).
\bb{dt}R. Hobart, Proc. Phys. Soc. Lond. {\bf82}, 201 (1963).
\bb{dt1}G.H. Derrick, J. Math. Phys. {\bf5}, 1252 (1964).

\bb{do}F. Doddato and J. MacDonald, JCAP {\bf1206}, 031 (1012); {\bf1307}, 004 (2013).
\bb{bar}S. Kasuya and M. Kawasaki, Phys. Rev. D {\bf89}, 103534 (2014). 
\bb{qbrane}S. Abel and A. Kehagias, {\it Q-branes}, arXiv:1507.04557.

\end{thebibliography}
\end{document}